\documentclass[journal,comsoc]{IEEEtran}

\ifCLASSINFOpdf

\else

\fi

\usepackage[utf8]{inputenc}
\usepackage{amssymb}
\usepackage{stfloats}
\usepackage{placeins}
\usepackage{physics}
\usepackage{overpic}
\usepackage{caption}
\usepackage{amsmath}
\usepackage{comment}
\usepackage{mathtools}
\usepackage{booktabs}
\usepackage{subfig}
\usepackage{lipsum}
\usepackage{graphicx}
\usepackage{textcomp,gensymb}
\usepackage{glossaries}
\usepackage{multicol}
\usepackage{algorithm}
\usepackage[noend]{algpseudocode}
\usepackage{tikz}
\usepackage[english]{babel}
\usepackage{amsthm}
\theoremstyle{plain}
\usepackage{dirtytalk}

\usepackage{enumitem}

\newtheorem{definition}{Definition}
\graphicspath{ {./images/} }

\usepackage{amsmath}

\usepackage{mathtools}

\usepackage{tabularx}
\newcolumntype{L}[1]{>{\raggedright\arraybackslash}p{#1}}
\newcolumntype{C}[1]{>{\centering\arraybackslash}p{#1}}
\newcolumntype{R}[1]{>{\raggedleft\arraybackslash}p{#1}}
\usepackage{multirow}
\usepackage{mathtools, cuted}

\newtheorem*{remark}{Remark}
\addto\captionsenglish{}
\newacronym{ts}{TS}{time sensitive}
\newacronym{aoi}{AoI}{age-of-information}
\newacronym{voi}{VoI}{value-of-information}
\newacronym{qaoi}{QAoI}{query age-of-information}
\newacronym{bs}{BS}{base station}
\newacronym{harq}{HARQ}{hybrid automated repeat request}
\newacronym{ee}{EE}{energy efficiency}
\newacronym{adc}{ADC}{analog-to-digital converter}
\newacronym{urllc}{URLLC}{ultra reliable low latency communication}
\newacronym{iot}{IoT}{internet of things}
\newacronym{pdf}{PDF}{probability density function}
\newacronym{cdf}{CDF}{cumulative distribution function}
\newacronym{kpi}{KPI}{key performance indicator}
\newacronym{snr}{SNR}{signal-to-noise ratio}
\newacronym{arq}{ARQ}{automatic repeat request}
\newacronym{iid}{i.i.d}{independent and identically distributed}
\newacronym{5g}{$\textrm{5}$G}{fifth-generation}
\newacronym{6g}{$\textrm{6}$G}{sixth-generation}
\newacronym{cpu}{CPU}{central processing unit}
\newacronym{wsn}{WSN}{wireless sensor network}
\newacronym{spa}{SPA}{Saddle Point Approximation}
\newacronym{goc}{GoC}{Goal-oriented Communication}
\newacronym{cscg}{CSCG}{circularly symmetric complex Gaussian}
\newacronym{twi}{TWI}{temporal window of integration}
\newacronym{ss}{SS}{synchronisation signal}
\newacronym{sr}{SR}{scheduling request}
\newacronym{pt}{PT}{packet transmission}
\newacronym{rr}{RR}{receiver response}
\newacronym{ptp}{PTP}{precision time protocol}
\newacronym{pmf}{PMF}{probability mass function}
\newacronym{gp}{GP}{geometric programming}
\newacronym{mgf}{MGF}{moment generating function}
\newacronym{dcp}{DCP}{disciplined convex programming}
\newacronym{xr}{XR}{extended reality}
\newacronym{ai}{AI}{artifical intelligence}
\newacronym{isac}{ISAC}{integrated sensing and communications}
\newacronym{hlc}{HLC}{hybrid logical clocks}
\newacronym{d2d}{D2D}{device-to-device}
\newacronym{d2d-ack}{D2D-ACK}{D2D acknowledgement}

\usepackage[backend=biber, style=ieee, maxnames=2, minnames=1, doi=false, url=false, isbn=false]{biblatex}
\usepackage{caption}
\usepackage{pgfplots}
\usepgfplotslibrary{fillbetween}

\definecolor{analytical}{HTML}{D95319}
\definecolor{simulated}{HTML}{bbbbbb}
\definecolor{twi_uniform}{HTML}{2b83ba}
\definecolor{twi_optimal}{HTML}{d7191c}
\definecolor{path1}{HTML}{d95f02}
\definecolor{path2}{HTML}{bbbbbb}

\pgfplotsset{
    compat=1.18,
    tick label style={font=\scriptsize},
    label style={font=\footnotesize},
    legend style={font=\scriptsize,draw=none,row sep=0,inner sep=0,outer sep=0,fill=none},
    tick style={color=black},
    major tick length=3pt,
    minor tick length=1.5pt,
    label shift=-4pt,
    pmfPlot/.style={
        ybar,
        bar width=0.5pt,
        enlarge x limits=0.05,
        width=0.5\columnwidth,
        height=0.45\columnwidth,
        xlabel={$n$},
        yticklabel style={/pgf/number format/fixed,/pgf/number format/zerofill,/pgf/number format/precision=1},
        grid=major,
        table/col sep=comma,
    },
    meanSubplot/.style={
        ybar=3pt,
        bar width=11pt,
        enlarge x limits=0.6,
        width=0.5\columnwidth,
        height=0.45\columnwidth,
        xlabel={\phantom{$n$}},
        ylabel={Mean (ms)},
        symbolic x coords={Mean,Variance}, xtick=data,
        nodes near coords style={font=\tiny,color=black,/pgf/number format/fixed,/pgf/number format/zerofill,/pgf/number format/precision=2},
        grid=major,
        axis y line*=left,
    },
    varianceSubplot/.style={
        ybar=3pt,
        bar width=11pt,
        enlarge x limits=0.6,
        width=0.5\columnwidth,
        height=0.45\columnwidth,
        ylabel={Variance (m$^2$s$^2$)},
        yticklabel style={/pgf/number format/fixed,/pgf/number format/zerofill,/pgf/number format/precision=1},
        symbolic x coords={Mean,Variance}, xtick=data,
        nodes near coords style={font=\tiny,color=black,/pgf/number format/fixed,/pgf/number format/zerofill,/pgf/number format/precision=2},
        legend pos=north east,legend columns=2,
        grid=major,
        axis y line*=right, axis x line=none,
    },
    twiBarPlot/.style={
        ybar=-11pt,
        bar width=11pt,
        enlarge x limits=0.55,
        width=0.45\columnwidth,
        height=0.45\columnwidth,
        xlabel={},
        ylabel={TWI (ms)},
        symbolic x coords={Optimal,Uniform}, xtick=data,
        nodes near coords*={$\approx$\pgfmathprintnumber{\pgfplotspointmeta}}, nodes near coords style={font=\tiny,color=black},
        grid=major,
    },
    reliabilityBarPlot/.style={
        ybar=6pt,
        bar width=11pt,
        enlarge x limits=0.6,
        width=0.55\columnwidth,
        height=0.45\columnwidth,
        xlabel={},
        ylabel={$\varepsilon_k$},
        yticklabel style={/pgf/number format/fixed,/pgf/number format/zerofill,/pgf/number format/precision=2},
        symbolic x coords={Optimal,Uniform}, xtick=data,
        nodes near coords*={$\approx$\pgfmathprintnumber{\pgfplotspointmeta}}, nodes near coords style={font=\tiny,color=black,/pgf/number format/fixed},
        legend pos=north east,legend columns=2,
        grid=major,
    },
    twiLinePlot/.style={
        width=0.5\columnwidth,
        height=0.45\columnwidth,
        grid=major,
    },
    optimalLine/.style={
        thick,
        twi_optimal,
        mark=*,
        mark size=1.1pt,
    },
    optimalArea/.style={
        twi_optimal,
        opacity=0.1,
    },
    uniformLine/.style={
        thick,
        twi_uniform,
        mark=square*,
        mark size=1.15pt,
    },
    uniformArea/.style={
        twi_uniform,
        opacity=0.1,
    },
}

\begin{document}

\title{{Temporal Windows of Integration for Multisensory Wireless Systems as Enablers of Physical AI}}

\author{\IEEEauthorblockN{ Anup~Mishra, \IEEEmembership{Member, IEEE},  João Henrique Inacio de Souza, \IEEEmembership{Member, IEEE}, Petar~Popovski, \IEEEmembership{Fellow, IEEE}}\vspace{-0.8cm}

\thanks{The authors Anup Mishra, João Henrique Inacio de Souza, and Petar Popovski are with the Connectivity Section, Department of Electronic Systems, Aalborg University, Aalborg 9220, Denmark (e-mail:anmi@es.aau.dk;
jhids@es.aau.dk;petarp@es.aau.dk).}}



\maketitle

\begin{abstract}
Physical \gls{ai} refers to the \gls{ai} that interacts with the physical world in real time. Similar to multisensory perception, Physical \gls{ai} makes decisions based on multimodal updates from  sensors and devices. With this, the Physical \gls{ai} acts as an artificial intelligent organism with a certain spatial footprint of its sensory tributaries. The multimodal updates traverse heterogeneous and unreliable paths, involving wireless links.
Throughput or latency guarantees do not ensure correct decision-making, as misaligned, misordered, or stale inputs still yield wrong inferences. Preserving decision-time coherence hinges on three timing primitives at the network–application interface: \emph{(i) simultaneity}, a short coincidence window that groups measurements as co-temporal, \emph{(ii) causality}, path-wise delivery that never lets a consequence precede its precursor, and \emph{(iii) usefulness}, a validity horizon that drops information too stale to influence the current action. In this work, we focus on the usefulness and we adopt \emph{\gls{twi}–Causality}: the \gls{twi} enforces decision-time usefulness by assuming path-wise causal consistency and cross-path simultaneity are handled upstream. We model end-to-end path delay as the sum of sensing/propagation, computation, and access/transmission latencies, and formulate network design as minimizing the validity horizon under a delivery reliability constraint. {In effect, this calibrates delay–reliability budgets for a timing-aware system operating over sensors within a finite spatial footprint.} The joint choice of horizon and per-path reliability is cast as a convex optimization problem, solved to global optimality to obtain the minimal horizon and per-path allocation of reliability. This is compared favourably to a benchmark based on uniform-after-threshold allocation. {Overall, this study contributes to  timing-aware Physical \gls{ai} in next-generation networks.}
\end{abstract}

\IEEEpeerreviewmaketitle
\glsresetall
\section{Introduction}
Perceptive wireless networks are poised to tightly integrate the capabilities of communication, sensing, localization, edge computing, and \gls{ai} in next-generation mobile systems~\cite{Khaled_Precpective,Simul_Causal,mishra2024coexistence_SR}. This integration will underpin {Physical \gls{ai}}, fusing multimodal observations from \glspl{bs}, \gls{iot} devices, and external sensors into actionable knowledge about the physical world~\cite{Khaled_Precpective,mishra2024coexistence,Simul_Causal}. Realizing this vision elevates \emph{timing} from a performance metric to a design primitive. As \gls{5g}’s latency targets broaden into \gls{6g}’s explicit timing guarantees, systems must reason about temporal ordering, simultaneity, and causality across heterogeneous links and modalities~\cite{popovski2022perspective,parvez2018survey,Simul_Causal,yates2021age,mishra2025reinforcementlearning}. Operationally, this shift translates into a run-time constraint: real-time decision-making requires inputs that are time-consistent and correctly ordered at the instant of action, despite inevitable network delays, clock impairments, and message loss or reordering~\cite{Wei_OOS,Hao_tracking,alkhateeb2023real,akyildiz2022wireless,ball2022metaverse}. These timing demands cut across how the network senses and transports information. Taken together, devices, sensors, and the network infrastructure form a multisensory perceptual system, {with the \gls{bs}/edge acting as an artificial brain that must fuse updates from sensors distributed over a finite region of the physical world}. The fidelity of this system hinges on precise temporal alignment across input streams~\cite{liu2022integrated,vroomen2010perception}. {Such} timing and alignment requirements defining a problem space comprising three key thrusts: $(i)$ characterising and enforcing simultaneity across distributed sensors~\cite{Joao_SV}, $(ii)$  maintaining causal consistency under network-induced asynchrony~\cite{Baldoni_DeltaCausal}, and $(iii)$ scheduling, filtering, and fusing multimodal streams so that inference remains resilient when observations arrive late, out of order, or not at all~\cite{Hao_tracking,Ahuja_MML}.
\begin{figure}[!t]
  \centering
  \subfloat[]{\includegraphics[width=210pt]{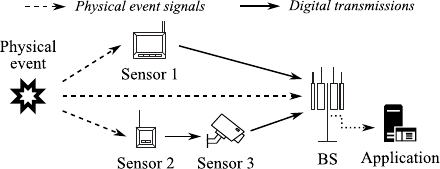}}\\\vspace{-8pt}
  \subfloat[]{\includegraphics[width=.90\columnwidth]{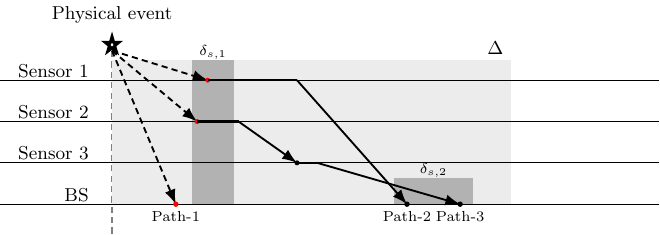}}\\\vspace{-6pt}
  \subfloat[]{\includegraphics[width=.99\columnwidth]{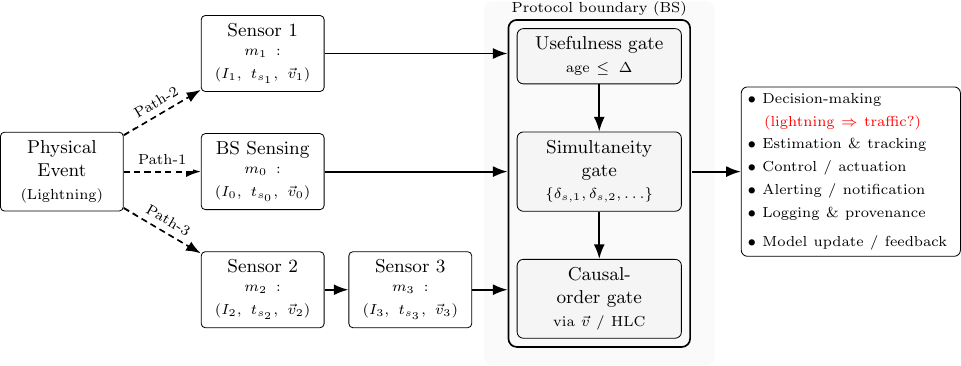}}
 \caption{(a) Toy multisensory scenario: a physical event (e.g. lightning) with three causal paths to the \gls{bs} (direct, via Sensor~1, via Sensor~2$\rightarrow$Sensor~3). 
(b) \gls{bs} timeline showing the validity horizon ($\Delta$) and coincidence windows $\delta_s$. 
(c) {Causal map and protocol-boundary processing (usefulness, simultaneity, causal order) prior to fusion/inference. Each message $m_k = (I_k, t_{s_k}, \vec v_k)$ carries content $I_k$, a timestamp $t_{s_k}$ taken when the signal first enters the digital domain at the corresponding sensor (and upon reception at relays and the \gls{bs}), and vector-clock metadata $\vec v_k$. In (b), red dots at dashed-arrow tips mark when signals first enter the digital domain and their generating events are timestamped.}}
\vspace{-0.6cm}
\label{fig:Toy_model}
\end{figure}
\par In asynchronous distributed systems, classical message ordering via logical clocks ensures consistent delivery order, but it is agnostic to physical-time constraints and cannot timestamp pre-digital physical events~\cite{lamport2019time,birman9lightweight,Simul_Causal} {since, obviously, \emph{nature does not put timestamps}}. Therefore, to operationalize timing in perceptive wireless networks with multimodal inputs, the above thrusts can be separated at the protocol boundary into {simultaneity}, {causality}, and {usefulness}.\footnote{We use {protocol boundary} to denote the interface at which the network stack hands updates to the \gls{bs}/edge application or fusion module.} {Simultaneity} is {based on device timestamps}: measurements are treated as co-temporal only if their timestamps fall within a small coincidence interval $\delta_s$~\cite{Joao_SV}, {akin to the `horizon of simultaneity' observed in human multisensory perception~\cite{vroomen2010perception}}. Causality can be enforced via logical clocks (e.g., vector clocks) or \gls{hlc}: both induce a happens-before partial order, with \gls{hlc} leveraging  timestamps opportunistically while falling back to logical counters to remain correct under clock impairments~\cite{Baldoni_DeltaCausal,Simul_Causal}. Usefulness is captured by a finite validity horizon $\Delta$: it is the maximum staleness the application tolerates for an update to remain actionable. Beyond $\Delta$, information is treated as too old to influence the current decision and is dropped rather than being delivered to the application~\cite{baldoni1996causal}. {Interestingly, usefulness generalizes the usual \gls{5g} \gls{urllc} link-level latency requirement: rather than a least-common-denominator bound for all services, it asks whether input data streams arrive before the application’s decision deadline. The validity horizon $\Delta$ captures this end-to-end usefulness budget, with the \gls{urllc} hop contributing a predictable slice~\cite{popovski2022perspective}.} Together, these choices yield a timing-aware delivery rule, referred to as $\Delta$-causal ordering in classical literature, where an input is eligible for application delivery only if it both respects causal order and arrives with age $\le \Delta$~\cite{baldoni1996causal,Baldoni_DeltaCausal,birman9lightweight}, thereby accommodating time validity and losses under unreliable, variable-delay paths~{\cite{baldoni1996causal,verissimo1994ordering,Marious@Usefulness,Marious@Filtering,Marious@QoS}}.
\par {Consider the toy example in Fig.~\ref{fig:Toy_model}(a) of a lightning strike} that reaches the \gls{bs} in three ways: the \gls{bs} senses it directly; Sensor~1 detects it and relays to the \gls{bs}; and Sensor~2 detects it, triggers a traffic camera (Sensor~3) to capture a traffic event, and that joint observation is forwarded to the \gls{bs}. {At each sensor (and at the \gls{bs} upon reception), the analog signal crosses into the digital domain and is timestamped; these instants are indicated in Fig.~\ref{fig:Toy_model} by explicit markers.} At the protocol boundary shown in Fig.~\ref{fig:Toy_model}(c), delivery passes through three timing gates. The usefulness gate in Fig.~\ref{fig:Toy_model}(c) enforces timeliness by admitting only paths whose age is within the validity horizon~$\Delta$; the light-gray region in Fig.~\ref{fig:Toy_model}(b) illustrates this $\Delta$-window.\footnote{{The} age $\Delta$ is \emph{designed} with respect to the (unobserved) true event time. {The actual} age of an update is \emph{estimated} using the best available proxy based on different observation timestamps. This proxy preserves the gating semantics; when it is conservative, it only tightens admission, not loosens it.} {Simultaneity} gate groups admitted paths that land within the same coincidence window on the \gls{bs} timeline, shown by the dark gray bands in Fig.~\ref{fig:Toy_model}\,(b). In practice, the coincidence window can first be used as an inference knob, relying on observation timestamps to adjudicate simultaneity directly, as in $\delta_{s,1}$. It can also serve as an engineering lever: the network can nudge selected paths to arrive within a common window, as in $\delta_{s,2}$, so that their updates are fused jointly and control deadlines are easier to meet. The causal-order gate delivers updates consistent with happens-before; vector/\gls{hlc} clocks infer that partial order. For example, in Path-$3$, a pair of vector clocks $\vec{v}_2=(1,0)$ and $\vec{v}_3=(1,1)$ indicates that Sensor~3’s update was triggered by Sensor~2, whereas $\vec{v}_2=(1,0)$ and $\vec{v}_3=(0,1)$ suggest that Sensor~3 reported independently. Timestamps ($t_{s_{1}},t_{s_{2}},t_{s_{3}}$) and coincidence windows complement this causal structure. Utilizing these three gates, the application at the \gls{bs}/edge determines actionability and either infers that the lightning caused the traffic event or treats the observations as coincidence. Similarly, it may trigger other actions, as shown in Fig.~\ref{fig:Toy_model}(c).
\par {Building on the above instantiation of the three timing primitives, in this work, we build a network-centric framework that operationalizes  timing in perceptive wireless networks.} Within this framework, we focus on usefulness at the protocol boundary, assuming cross-path simultaneity is handled by timestamping and per-path causality is enforced upstream through \glspl{hlc}. Building on this separation, we adopt \emph{\gls{twi}–Causality}, {an instantiation of \(\Delta\)-causal ordering tailored to perception in multisensory/multimodal scenarios. In \gls{twi}–Causality}, the \gls{twi} acts as the validity horizon \((\text{TWI}\equiv\Delta)\) that gates delivery. An update is eligible for hand-off to the application only if causal order holds and its arrival lies within \gls{twi} relative to the originating event; arrivals outside \gls{twi} are treated as temporally obsolete and discarded. In wireless pipelines spanning sensing, computation, random access, scheduling, and transport, this discipline yields bounded reordering buffers and deadline-constrained delivery. {This makes it explicit that there is no universal, assumption-free notion of temporal coherence: both human perception and engineered multisensory systems must define guarantees relative to a given spatial footprint, latency model, and reliability target.} {In our setting, this leads to the design problem of deriving the smallest \gls{twi} ($\Delta$) that preserves decision-time coherence at a target reliability across admissible causal paths~\cite{Simul_Causal,baldoni1996causal}.}
\subsection{Related Works}
The traditional logical-clock literature has explored causal and message ordering using vector clocks and related metadata, ensuring in-order delivery, {yet without grounding in} physical-time semantics~\cite{lamport2019time,birman1987reliable,birman9lightweight}. {Next}, \cite{baldoni1996causal,verissimo1994ordering} admitted a timing constraint at delivery: under $\Delta$-causal ordering, messages are released only when {a causal order holds and the age is not greater than $\Delta$.} {As logical ordering is} unsuitable for \glspl{wsn} with long, variable delays, \cite{kay2003temporal} examines temporal message ordering with emphasis on energy efficiency, scalability, and immediacy.
\par {Multisensory perception studies have explored} inter-sensory synchrony, aligning with engineering treatments of simultaneity~\cite{vroomen2010perception,spence2003multisensory}. {Research on distributed cyber-physical systems has related} chronology, simultaneity, and causality to clock synchronisation and network/computation latencies, extending consistency to encompass physical state~\cite{popovski2022perspective,yates2021age}. Motivated by human multisensory perception, \cite{Simul_Causal} {uses \gls{twi} to impose} timing constraints on processing and preserve temporal ordering while enabling the tracking of causal relations. 
\par From the wireless network perspective, \gls{isac}-oriented visions for \gls{6g} cast \glspl{bs}/ edge processors as multisensory platforms that combine radio-native sensing with transparent sensing over digital links~\cite{liu2022integrated,Khaled_Precpective,Wild_ISAC}. Architectural extensions add passive target monitoring terminals and surface system issues such as clock synchronisation and time transfer across the infrastructure~\cite{Xie_TMT}. Separately, out-of-order arrivals have been examined through the \gls{aoi} lens, which emphasizes freshness metrics and timeliness~\cite{Pappas_AoI}. Within this infrastructure view, the human-multisensory perspective posited in~\cite{Simul_Causal} motivates deriving network-level \glspl{twi} and a composite latency model covering propagation, sensing, computation, and communication. Building on this foundation, \cite{Joao_SV} adopts an event-centric perspective in a two-sensor setting and uses the probability of simultaneity violation to size the coincidence window, balancing chronological preservation against event throughput.
\par {From a deployable timing perspective, and to the best of our knowledge, conventional literature is largely centred on message ordering, whereas the multisensory perception–oriented framework remains nascent and lacks a network-centric notion of timing primitives. This motivates a protocol-boundary notion of timeliness within the multisensory framework; specifying and sizing a \emph{network-level} validity horizon ($\Delta$/\gls{twi}) under variable delays along admissible causal paths to ensure decision-time usefulness, with simultaneity and ordering residing within the horizon.} {In practical terms, a network-level validity horizon enables deploying inference engines with a clear prescription of how far their sensory tributaries may extend and how long they must integrate updates to maintain decision-time coherence at a given reliability. Rather than pursuing a universal notion of temporal consistency, the framework makes explicit the spatial footprint {of the Physical AI}, timing budgets, and reliability targets under which temporal coherence is engineered.}
\vspace{-0.1cm}
\subsection{Contributions {and Paper Organization}}
Building on perception-inspired timing abstractions in~\cite{Simul_Causal,Joao_SV}, this study takes a network-centric view and analyzes the protocol-boundary validity horizon $\Delta$ (\gls{twi}) for multi-sensor, multi-path pipelines under variable path delays. To this end, this work addresses \gls{twi} design for temporal coherence in a wireless sensing pipeline where an event triggers multiple causal paths (possibly multi-hop) delivering multimodal updates to a \gls{bs}. As noted above, causality within each admissible path is enforced by \glspl{hlc}, while cross-path simultaneity is handled by timestamping. Our main contributions are:

\begin{enumerate}
\item We consider a general cellular system model, define causal paths, and build a tractable end-to-end latency model per causal path. {We assume that sensor nodes are synchronised to a common time reference within a known accuracy and timestamp each acquisition as it enters the digital domain; relays and the \gls{bs} also timestamp packet receptions.} The latency path explicitly composes event–to–sensor propagation, bounded computation at each hop, and protocol-induced communication delays with slotted frames and retry limits. Multi-hop sensor relays are captured via truncated-attempt models (grant-free for intermediate hops, grant-based for the final hop), yielding closed-form packet-drop probabilities and enforcing alignment to frame boundaries.

\item {Timeliness and reliability guarantees are computed from the \glspl{pmf} and their moments, derived from analysis of heterogeneous contributors (propagation, computation, protocol delays) within a causal path.}

\item We pose a drop-aware reliability requirement with respect to the physical event time: find the smallest integration window (validity horizon $\Delta$) such that each admissible causal path delivers within that horizon with high reliability. We distribute a global reliability budget across paths via per-path violation budgets. Using only the first two moments, we then derive a distribution-agnostic lower bound via the Cantelli (one-sided Chebyshev) margin; the tightest one-sided guarantee available with first two moments absent additional structural assumptions~\cite{Tight_Cantelli}. The resulting constraints admit a convex optimization formulation, yielding a globally optimal \gls{twi} under the specified reliability target.

\item We jointly optimize the per-path violation budgets and the \gls{twi}, benchmarking against a uniform-after-threshold baseline that admits a water-filling allocation to isolate the gains from reliability shaping. Through numerical results, this comparison quantifies the improvement achieved by targeted (non-uniform) reliability allocations. It should be highlighted that the optimized per-path reliabilities 
can serve as availability signals for reliability-aware weighting and scheduling in fusion and inference, and can also provide auditable per-path confidence for temporal forensics, enabling bias-aware analyses under differential timeliness~\cite{ZHANG2025110663,jones2005health,Fitzerald_IPW}. Furthermore, we quantify how network parameters such as cell size, frame time, access periodicity, causal-path width, path depth, and the global reliability target, shape both the optimal allocation and the \gls{twi}. These sensitivities translate into design guidance on the system parameters that dominate timeliness in this perception-inspired timing-abstractions framework.
\end{enumerate}
{Taken together, these contributions provide a network-centric design methodology for timing-aware systems such as Physical \gls{ai}, in which an inference engine at the \gls{bs}/edge is co-designed with its spatio-temporal sensing envelope (e.g., a radius-$D$ sensor field) and reliability requirements.}

{\emph{Organisation:}} Section \ref{Sys_Mod} defines admissible causal paths and enumerates their latency constituents. Section \ref{Prob_Form} formulates the problem of finding the minimal \gls{twi} under a global reliability target via per-path violation budgets. Section \ref{opt_frame} derives \glspl{pmf} and moments for general causal paths, introduces the Cantelli (one-sided Chebyshev) margin to obtain distribution-agnostic constraints, proves convexity, and presents a solution via bisection with a convex-feasibility check. Section \ref{Num_Results} reports numerical results: we benchmark against a uniform-after-threshold (water-filling) baseline, quantify gains from optimal reliability shaping, and conduct sensitivity analyses over cell size, frame duration, causal-path width/depth, and the reliability target, distilling design implications. Finally, Section \ref{Conc} concludes and outlines directions for future work.

{\emph{Notation:}} Matrices are denoted by boldface uppercase letters, column vectors are denoted by boldface lowercase letters, and scalars are denoted by standard letters. 
{The expectation of $Y$ with respect to a random variable $X$ is denoted by $\mathbb{E}_{X}[Y]$ and the probability of an event $E$ by $\Pr(E)$.} {The sets} $\mathbb{C}^{M\times N}$ and $\mathbb{R}^{M\times N}$ denote 
all $M \times N$ matrices with complex- and real-valued entries, respectively. The \gls{cscg} distribution with mean $\mu$ and variance $\sigma^{2}$ is denoted by $\mathcal{CN}(\mu,\sigma^{2})$. {The uniform distribution over the interval $[a, b]$ is denoted by $\mathcal{U}[a, b]$. The geometric distribution with success probability $p$ is denoted by $\text{Geo}(p)$.} The Big-O notation $\mathcal{O}(f(n))$ describes the asymptotic upper bound of an algorithm's computational complexity. The floor and ceiling functions are denoted by $\lfloor x \rfloor$ and $\lceil x \rceil$, representing the greatest integer less than or equal to $x$ and the smallest integer greater than or equal to $x$, respectively.
\vspace{-0.1cm}
\section{System Model}\label{Sys_Mod}
We consider a \gls{wsn} deployed within a circular cell of radius $D$, with a \gls{bs} at the centre. Sensors are distributed uniformly at random over the cell. The sensor population is heterogeneous with two functional classes:
\begin{enumerate}
    \item \emph{BS-reachable sensors} (e.g., aggregators/edge nodes) that can sense, perform local computation, and communicate {directly} with the \gls{bs};
    \item \emph{Relay sensors} (low-power nodes) that can sense and process but communicate only via short-range \gls{d2d}, forwarding through \gls{bs}-reachable sensors.
\end{enumerate}
Upon the occurrence of a physical event at time \( t_0=0 \), status updates may reach the \gls{bs} through the following mechanisms:
\begin{itemize}
    \item[(a)] The event is directly sensed by the \gls{bs};
    \item[(b)] A \gls{bs}-reachable sensor detects the event and transmits its update directly to the \gls{bs};
    \item[(c)] A relay sensor detects the event and forwards its update to a nearby \gls{bs}-reachable sensor, which in turn relays it to the \gls{bs}.
\end{itemize}
In addition to simple forwarding, causal detection may also initiate {cross-sensor triggering}, where a detecting sensor prompts other proximate or semantically relevant sensors (e.g., with complementary modalities or measurement functions) to upload their data to the \gls{bs}. In cases (b) and (c){, as seen in Fig.~\ref{fig:communication-frame},} these event-driven updates may traverse multiple wireless hops, forming what we refer to as a \emph{causal path}: an ordered sequence of sensor-to-sensor and/or sensor-to-\gls{bs} transmissions initiated by a single event detection. Following this, we proceed to defining a causal path.
\begin{definition}
A {causal path} \( \pi_k \) is an ordered sequence of nodes through which an event-triggered status update propagates to the \gls{bs}:
\begin{equation}
\pi_k = \{ s_{k,1} \rightarrow s_{k,2} \rightarrow \cdots \rightarrow s_{k,h_k} \rightarrow \text{\gls{bs}} \},
\end{equation}
where \( s_{k,i} \) is the $i^{\text{th}}$ sensor on the $k^{\text{th}}$ path, \( h_{k} \in \mathbb{N}_0 \) denotes the number of sensor nodes on that path, and $k\in\{1,\ldots,K\}$. For the special case of direct \gls{bs} sensing, the path reduces to the notation \( \pi_k = \{\text{BS}\} \). The arrow \( \rightarrow \) represents a directed communication link.
\end{definition}
Let \( \mathcal{P} = \{ \pi_1, \pi_2, \dots, \pi_K \} \) denote the set of all causal paths triggered by a given event. Each path \( \pi_k \in \mathcal{P} \) is defined as an ordered sequence of nodes through which an event-triggered status update propagates to the \gls{bs}.\footnote{We restrict attention to \emph{non-merging causal paths}, i.e., no two paths are allowed to intersect at intermediate nodes before reaching the \gls{bs}. This assumption simplifies independence across path delays and avoids modeling correlation and queuing effects that arise from shared relay nodes. The more general case of merging paths is left for future work.} The total delay associated with this path, denoted as \( T_{\pi_k} \), is a random variable comprising multiple latency components introduced along the path. These include sensing delays, computation times at intermediate nodes, and communication delays between nodes and toward the \gls{bs}, each influenced by physical distance, access protocols, and channel conditions. To explicitly characterize the random delay \( T_{\pi_k} \) incurred along a causal path, we decompose it into the following constituent components:
\begin{itemize}[leftmargin=*]
    \item \textbf{Event propagation delay}: The time taken for the physical signal to reach the first node on the causal path depends on the relative position of the event source and the receiving node. The node could be the \gls{bs} or a sensor.
    \begin{itemize}
         \item \emph{Event-to-\gls{bs}}: The \gls{bs} is located at the center of the cell (disk of radius $D$), while the event occurs at a uniformly random position within the disk. Let \( R_{\text{BS}} \) be the distance from the event to the BS. Subsequently, its \gls{pdf} is given by~\cite{Joao_SV}:
         \begin{equation}
            f_{R_{\text{BS}}}(r) = \frac{2r}{D^2}, \quad r \in [0, D].
         \end{equation}
         Then, assuming the propagation delay to the \gls{bs}, $T_{\text{BS}}^{\text{prop}}$, is proportional to $R_{\text{BS}}$, we have:
        \begin{equation}
            T^{\text{prop}}_{\text{BS}} = \frac{R_{\text{BS}}}{v},
        \end{equation}
       where $v$ is the propagation speed of the physical signal. The \gls{pdf} of this propagation delay is obtained using change of variables as:
        \begin{equation}
            f_{T^{\text{prop}}_{\text{BS}}}(t) = \frac{2v^2 t}{D^2}, \quad t \in \left[0, \frac{D}{v} \right].
        \end{equation}
    
        \item \emph{Event-to-sensor}: Let \( R_{\text{ES}} \) denote the Euclidean distance between the event source and the first sensor on a causal path. When both points are independently and uniformly distributed in a disk of radius \( D \), the \gls{pdf} of \( R_{\text{ES}} \) is expressed as~\cite{mathai1999introduction}:
        \begin{equation}
            f_{R_{\text{ES}}}(r) = \frac{4r}{\pi D^2} \left[ \cos^{-1}\left( r_{D} \right) - r_{D} \sqrt{1 - \left( r_{D} \right)^2} \right],
        \end{equation}
        where $r_{D}=r/2D$, and $r \in [0, 2D]$. Assuming $T_{\text{ES}}^{\text{prop}}$ is proportional to $R_{\text{ES}}$, we define:
        \begin{equation}
            T_{\text{ES}}^{\text{prop}} = \frac{R_{\text{ES}}}{v}.
        \end{equation}
        Applying a change of variable with \( R_{\text{ES}} = v \cdot T_{\text{ES}}^{\text{prop}} \), the PDF of \( T_{\text{ES}}^{\text{prop}} \) becomes:
        \begin{equation}\label{eq:ES_Prop_time}
            f_{T_{\text{ES}}^{\text{prop}}}(t) = \frac{4v^2 t}{\pi D^2} \left[ \cos^{-1}\left( v_{t} \right) - v_{t} \sqrt{1 - \left( v_{t} \right)^2} \right],
        \end{equation}
        for $v_{t}=\frac{v\,t}{2D} $, and \( t \in \left[ 0, \frac{2D}{v} \right] \). This model captures the randomness of the initial signal propagation from a randomly located event source to a randomly positioned sensor within a cell of radius \( D \).
    \end{itemize}
    \item \textbf{Computation Delay:}
    Each sensor incurs a local computation delay before transmitting or relaying a status update. For the first sensor on a causal path, this is the time to process the event signal and generate the update. For intermediate nodes, it reflects processing that fuses complementary modality readings with the incoming information before encoding and transmitting to the next node. We model the computation delay at node \( i \) in causal path $\pi_{k}$, denoted by \( C_{k,{i}} \), as a bounded uniform random variable~\cite{Simul_Causal}:
    \begin{equation}\label{eq:Comp_time}
    C_{k,i} \sim \mathcal{U}[C_{\min}, C_{\max}].
    \end{equation}
    This reflects simple embedded processing with bounded uncertainty, and supports tractable analysis when combined with other random components~\cite{Joao_SV}.
    \begin{figure*}[!t]
        \centering
        \includegraphics[width=492.48pt]{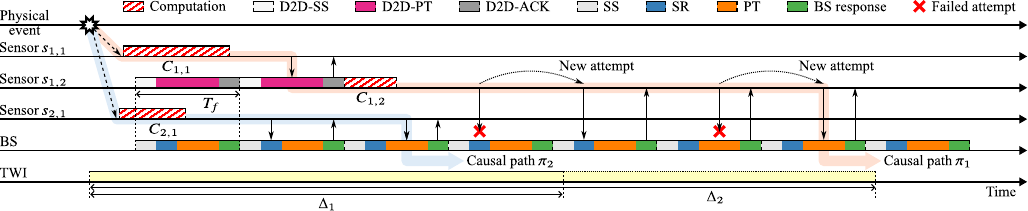}
        \caption{{Timing diagram of computation and communication delays for two causal paths: $\pi_1$ with two sensors ($s_{1,1},s_{1,2}$) and $\pi_2$ with one sensor ($s_{2,1}$). The paths experience different delays due to hop count and retransmissions along $\pi_1$. The task is to design \gls{twi} ($\Delta$) so that, with reliability $1-\varepsilon$, all admissible paths deliver within $\Delta$. The markers $\Delta_1$ and $\Delta_2$ indicate candidate horizons. In this instance, with $\Delta_1$ as the \gls{twi}, the \gls{bs} excludes $\pi_1$; with $\Delta_2$, both paths are admitted, albeit with higher decision latency.}}\vspace{-0.3cm}
       \label{fig:communication-frame}
    \end{figure*}
    \item \textbf{Communication Delay}: We consider a frame-based wireless communication system with frame duration \( T_f > 0 \), comprising downlink and uplink sub-frames. {In sensor-to-\gls{bs} communication,} each frame includes \gls{ss}, \gls{sr}, \gls{pt}, and \gls{rr} {subframes}~\cite{Joao_SV}. {In contrast, in sensor-to-sensor communication, each frame includes \gls{d2d}-\gls{ss}, \gls{d2d}-\gls{pt}, and \gls{d2d-ack} subframes.} {We assume all nodes are synchronised to a common time base using \gls{ptp}, enabling slotted transmissions and coordinated access across the network~\cite{PTP_Cho}. Consequently, the sensor–\gls{bs} and sensor–sensor frames are taken to have aligned boundaries (see Fig.~\ref{fig:communication-frame}).\footnote{The impact of residual synchronisation errors, drift, or jitter on multi-hop causal-path delay and reliability is left for future investigation. These effects predominantly influence causal order and simultaneity.}}  Based on this frame structure, we differentiate two types of communication in a causal path $\pi_{k}\in \mathcal{P}$:
    \begin{itemize}
        \item \textit{Sensor-to-BS Communication:} This communication follows a \emph{grant-based} protocol. A sensor first transmits a unique orthogonal \gls{sr} preamble during the \gls{sr} sub-frame. Upon successful detection, the \gls{bs} allocates a scheduling grant, and the sensor transmits its status update in the next frame~\cite{Joao_SV}. Communication delay is modelled as two separate geometric components: one for \gls{sr} access and one for \gls{pt}. To analyse this delay, we evaluate the \gls{pmf} of the total time needed for both \gls{sr} and \gls{pt}. The probability that the \gls{bs} fails to detect the \gls{sr} from sensor~\( i \) of causal path $\pi_{k}$ in a given frame is denoted by \( \zeta_{k,i} \), which depends on the uplink \gls{snr} \( \gamma_{k,i} \), and is modelled as~\cite{talli2025saving,Anders@timing,Joao_SV}:
        \begin{equation}
        \zeta_{k,i} = 1 - \left(1 + \gamma_{k,i} \right)^{- \frac{1}{\gamma_{k,i}}}.
        \end{equation}
        Assuming block-fading and independent access attempts across frames, the number of \gls{sr} attempts \( m_{k,i} \), with retry limits imposed, follows a truncated geometric distribution with success probability \( 1 - \zeta_{k,i} \)~\cite{Yang_FBL,Kay_ST}. The \gls{sr} delay is then given by:
        \begin{equation}
        T^{\text{\gls{sr}}}_{k,i} = m_{k,i} T_f, \quad \forall m_{k,i} \in \{1, 2, \dots, M_{\max}\},
        \end{equation}
        and
        \begin{equation}
        \Pr \left( T^{\text{\gls{sr}}}_{k,i} = m T_f \right) = \frac{\zeta_{k,i}^{m-1} (1 - \zeta_{k,i})}{1 - \zeta_{k,i}^{M_{\max}}}, 
        \end{equation}
        with $m = \{1, 2, \dots, M_{\max}\}$. Once a grant is received, the sensor proceeds with packet transmission. Subsequently, for \gls{pt} as well, assuming transmission success probability \( 1 - \epsilon_{k,i} \), the number of packet attempts \( n_{k,i} \) also follows a truncated geometric distribution. The transmission delay is expressed as:
        \begin{equation}
        T^{\text{\gls{pt}}}_{k,i} = n_{k,i} T_f, \quad \forall n_{k,i} \in \{1, 2, \dots, N_{\max}\},
        \end{equation}
        and the associated \gls{pmf} is:
        \begin{equation}
        \Pr \left( T^{\text{PT}}_{k,i} = n T_f \right) = \frac{\epsilon_{k,i}^{n-1} (1 - \epsilon_{k,i})}{1 - \epsilon_{k,i}^{N_{\max}}},
        \end{equation}
        with $n = \{1, 2, \dots, N_{\max}\}$. The total delay for sensor-to-BS communication is $T^{\text{SR}}_{k,i} + T^{\text{PT}}_{k,i}.$
        \item \textit{Sensor-to-Sensor Communication:} Modeled as a \emph{grant-free} one-shot transmission attempt per frame. Each hop between sensors entails a joint delay component, capturing both contention resolution and packet delivery, modeled via a geometric distribution over frame intervals. Let \( \rho_{k,i} \in [0,1) \) be the failure probability for a transmission from sensor \( i \) in  $\pi_k$ to its next-hop peer. Then, the number of attempts required before successful reception is a geometrically distributed random variable \( a_{k,i} \sim \text{Geo}(1 - \rho_{k,i}) \), with the resulting delay~\cite{Anders@timing,talli2025saving}:
        \begin{equation}
        T^{\text{hop}}_{k,i} = a_{k,i}T_f, \quad \forall a_{k,i} \in \{1, 2, \dots, A_{\max}\}.
        \end{equation}
        Similar to sensor-to-\gls{bs} communication, we impose a retry limit \( A_{\max} \in \mathbb{N} \). The \gls{pmf} of the sensor-to-sensor communication delay is then given by:
        \begin{equation}\label{eq:T_hop_aki}
        \Pr\left( T^{\text{hop}}_{k,i}\,\big|\, 1 \leq a \leq A_{\max} \right)
        =\frac{ \rho_{k,i}^{a-1} (1 - \rho_{k,i}) }{ 1 - \rho_{k,i}^{A_{\max}}}, 
        \end{equation}
        with $ a \in \{1, 2, \dots, A_{\max}\}$.
    \end{itemize}
    \item \textbf{Frame Quantization:} All communication processes are constrained to discrete frame boundaries. Accordingly, all delay components are aligned using ceiling operations of the form: $\left\lceil \frac{t}{T_f} \right\rceil T_f$, which capture the protocol-level quantization of timing and scheduling granularity.
\end{itemize}

Based on the exposition of delay components and constraint of quantisation, the total delay of a path \( \pi_k \) consisting of \( h_{k} \) sensors is:
\begin{equation}\label{eq:event_to_sensor_to_BS}
T_{\pi_k} = \left\lceil \frac{T_{\text{ES}}^{\text{prop}} + C_1}{T_f} \right\rceil T_f 
+ \sum_{i=1}^{h_k - 1} \left\lceil \frac{T^{\text{hop}}_{k,i} + C_{k,i}}{T_f} \right\rceil T_f
+ T^{\text{SR}}_{h_k} + T^{\text{PT}}_{h_k}.
\end{equation}
If $\pi_{k}$ is a direct path to the \gls{bs}, the event is sensed and reported by the \gls{bs} itself, then the total delay reduces to the event propagation delay to the \gls{bs}:
\begin{equation}
T_{\pi_k} = T_{\text{BS}}^{\text{prop}}.
\end{equation}
We neglect any computation delay at the \gls{bs}, assuming it has ample processing resources to handle incoming updates with negligible latency. Having quantified the total path delay, we proceed to the problem formulation.
\vspace{-0.3cm}
\section{Problem Formulation}\label{Prob_Form}
In this section, given a set of admissible causal paths to the \gls{bs}, we seek the smallest validity horizon (\gls{twi}, \(\Delta\)) such that, at a target network-level reliability \(1-\varepsilon\), every path’s update arrives within \gls{twi} under a non-uniform distribution of per-path violation budgets \(\{\varepsilon_k\}\). The resulting \(\text{TWI}^\star\) and \(\{\varepsilon_k^\star\}\) quantify the minimal horizon and the reliability profile across paths. Note that these path-level reliabilities are actionable at the protocol boundary of Fig.~\ref{fig:Toy_model}(c). On one hand, they inform fusion and inference via reliability-aware weighting and scheduling~\cite{ZHANG2025110663}. On the other hand, they support temporal forensics by accounting for the fact that some paths are timely more often than others; in practice, such imbalance can be corrected by re-weighting the data, modelling missing/late cases, or analysing groups by timeliness so conclusions are not skewed by which paths happened to be available. {This approach aligns with common practice in causal-inference studies across health, econometrics, and transportation, which regularly employ re-weighting and missing-data modelling~\cite{jones2005health,Fitzerald_IPW,ANUPRIYA2025108168}.}

\subsection{Network-level reliability via union bound.}
With $\mathcal{P}=\{\pi_1,\dots,\pi_K\}$ as the set of causal paths triggered by a single event, let $E_k$ denote the event that path $\pi_k$ {both} (i) successfully delivers an update (no packet drop) {and} (ii) the delivery delay does not exceed \gls{twi}. We require that, with high probability, \emph{all} paths meet these conditions simultaneously:
\begin{equation}\label{eq:Overall_R}
    \Pr\!\left(\bigcap_{k=1}^K E_k \right) \ge 1-\varepsilon,
\end{equation}
where $\varepsilon \ll 1$ determines overall system-level success and timeliness requirement. Next, applying the union bound, we transform the network-level reliability constraint to per-path constraint as~\cite{pishro2014introduction},
\begin{equation}
\label{eq:union-first}
    \Pr\!\left(\bigcap_{k=1}^K E_k \right)
    \;\ge\; 1 - \sum_{k=1}^K \Pr(E_k^c)
    \;\ge\; 1 - \sum_{k=1}^K \varepsilon_k
    \;\ge\; 1-\varepsilon,
\end{equation}
where we allocate per-path violation budgets $\{\varepsilon_k\}$ satisfying $\sum_k \varepsilon_k \le \varepsilon$ for tractability.
\subsection{Path-level reliability with packet drops.}
For each path $\pi_k$, define the {path drop probability} $P^{\text{drop}}_{\pi_k}$ as the probability that the path fails to deliver an update due to exhausting retry limits at one or more communication stages. Under our model (independent per-stage attempts), a tractable and conservative expression is:
\begin{equation}
\label{eq:drop-path}
    P^{\text{drop}}_{\pi_k} \;=\; 1
    \;-\;
    \left[
      \prod_{i=1}^{h_k-1} \big(1-\rho_{k,i}^{A_{\max}}\big)
    \right]
    \big(1-\zeta_{h_k}^{M_{\max}}\big)\,
    \big(1-\epsilon_{h_k}^{N_{\max}}\big),
\end{equation}
where each factor is the success probability of a stage within its retry budget:
(i) sensor-to-sensor hop succeeds within $A_{\max}$ attempts with probability $1-\rho_{k,i}^{A_{\max}}$,
(ii) \gls{sr} succeeds within $M_{\max}$ attempts with probability $1-\zeta_{h_k}^{M_{\max}}$,
(iii) \gls{pt} succeeds within $N_{\max}$ attempts with probability $1-\epsilon_{h_k}^{N_{\max}}$.
Let $F_{\pi_k}(t\,|\,\text{no drop}) \triangleq \Pr(T_{\pi_k}\le t\,|\,\text{no drop})$ denote the conditional \gls{cdf} of the end-to-end path delay given success (no drop). Then the {overall} path-level success-and-timeliness probability satisfies:
\begin{align}\label{eq:overall-path-success}
    \Pr(E_k)
    \;&=\;
    \Pr(\text{no drop})\cdot \Pr\,(T_{\pi_k}\le \text{TWI}\,|\,\text{no drop})\\
    \;&=\;
    \big(1-P^{\text{drop}}_{\pi_k}\big)\,F_{\pi_k}(\text{TWI}\,|\,\text{no drop}).
\end{align}
To be consistent with the per-path budgets in~\eqref{eq:union-first}, we enforce:
\begin{equation}
\label{eq:path-constraint}
    \big(1-P^{\text{drop}}_{\pi_k}\big)\,F_{\pi_k}(\text{TWI}\,|\,\text{no drop})
    \;\ge\;
    1-\varepsilon_k,
    \, \forall\,\pi_k\in\mathcal{P}.
\end{equation}
Equivalently, we express
\begin{equation}
\label{eq:conditional-tighten}
    F_{\pi_k}(\text{TWI}\,|\,\text{no drop})
    \;\ge\;
    \frac{1-\varepsilon_k}{1-P^{\text{drop}}_{\pi_k}}.
\end{equation}
\begin{remark}
Our formulation adopts an {absolute} notion of the \gls{twi}, defined relative to the time of occurrence of the physical event, despite this quantity being unobservable at runtime. This design choice is essential to preserve the semantics of \emph{\(\Delta\)-causality}, which stipulates that causally related updates must arrive within a finite validity window. Specifically, we impose that all causal paths \( \pi_k \in \mathcal{P} \) must satisfy the constraint \( T_{\pi_k} \leq \text{\gls{twi}} \) with high reliability, where \( T_{\pi_k} \) denotes the total end-to-end delay of path \( \pi_k \) measured from the time of occurrence of the event. This constraint ensures that the system supports reliable temporal fusion of multimodal information triggered by a single event. While a relative \gls{twi} (e.g., triggered by the first arrival) may be operationally simpler, it lacks causal grounding and degenerates to a packet-level delay bound~\cite{Joao_SV}. 
\end{remark}
\subsection{Optimization problem.}
Given the stochastic delay profiles of each path and a target global reliability $1-\varepsilon$, we seek the smallest \gls{twi} meeting~\eqref{eq:union-first} via the per-path constraints~\eqref{eq:path-constraint}:
\begin{subequations}\label{eq:opt-twi-drop}
\begin{align}
\underset{\text{TWI},\,\{\varepsilon_k\}}{\text{minimise}}\quad&
\text{TWI} \\
\text{s.t.}\quad
& \eqref{eq:conditional-tighten},
\quad \forall\pi_k\in\mathcal{P}, \label{eq:opt-twi-drop-a}\\
& \sum_{k=1}^K \varepsilon_k \le \varepsilon, \label{eq:opt-twi-drop-b}\\
& 0 \le \varepsilon_k \le 1, \quad \forall k \\
& 0 \le \varepsilon \le 1. \label{eq:opt-twi-drop-c}
\end{align}
\end{subequations}
If a direct event-to-\gls{bs} path exists, it can be included as $\pi_{\text{BS}}\in\mathcal{P}$ with $P^{\text{drop}}_{\pi_{\text{BS}}}=0$ and reliability target $\varepsilon_{\text{BS}}=0$, anchoring the lower bound on \gls{twi}. If no direct path is available, $\pi_{\text{BS}}$ is excluded and the feasible \gls{twi} is determined solely by multi-hop paths. The drop term $P^{\text{drop}}_{\pi_k}$ is closed-form via~\eqref{eq:drop-path}. The conditional \gls{cdf} $F_{\pi_k}(\cdot\,|\,\text{no drop})$ is induced by the composition in~\eqref{eq:event_to_sensor_to_BS}. 
\section{Optimisation Framework}\label{opt_frame}
\begin{figure}[t]
    \centering


    \subfloat[\emph{Light:} $N_1$ PMF]{        
        \begin{tikzpicture}
        \begin{axis}[
            pmfPlot,
            ylabel={$N_1$ PMF},
            xmin={1}, xmax={51}, xtick={1,10,20,30,40,50},
            ymin={0}, ymax={0.025}, ytick distance={0.005},
            legend columns=2, legend style={at={(0.5,1.1)}, anchor=south, /tikz/every even column/.append style={column sep=1em}},
        ]
            \addplot [only marks,analytical,mark size=0.75pt,forget plot] table [x=n,y=PMF_Analytical] {data/N1_PMF_results_Light.csv};
            \addlegendimage{line legend,only marks,mark size=1.5pt,color=analytical}
            \addlegendentry{Analytical}

            \addplot [fill=simulated,draw=simulated] table [x=n,y=PMF_Simulation] {data/N1_PMF_results_Light.csv};
            \addlegendentry{Simulated}
        \end{axis}
        \end{tikzpicture}
    }
    \subfloat[\emph{Light:} $\mathbb{E}\,\lbrack T_1\rbrack$ and $\mathrm{Var}(T_{1})$]{
        \begin{tikzpicture}
        \pgfplotsset{set layers}
        \begin{axis}[
            meanSubplot,
            ymin={200}, ymax={300}, ytick distance={25},
            legend columns=2, legend style={at={(0.5,1.1)}, anchor=south, /tikz/every even column/.append style={column sep=1em}},
        ]
            \addplot [fill=analytical,draw=analytical!400,nodes near coords*={$\approx$\pgfmathprintnumber{\pgfplotspointmeta}}] coordinates {
                (Mean,260.0003018) (Variance,-999)
            };
            \addlegendentry{Analytical}
            \addplot [fill=simulated,draw=simulated!300] coordinates {
                (Mean,260.01145) (Variance,-999)
            };
            \addlegendentry{Simulated}
        \end{axis}
        \begin{axis}[
            varianceSubplot,
            ymin={5000}, ymax={25000}, ytick distance={5000},
        ]
            \addplot [fill=analytical,draw=analytical!400,nodes near coords*={$\approx$\pgfmathprintnumber{\pgfplotspointmeta}}] coordinates {
                (Mean,-999) (Variance,20000.00302)
            };
            \addplot [fill=simulated,draw=simulated!300] coordinates {
                (Mean,-999) (Variance,20011.9538688975)
            };
        \end{axis}
        \end{tikzpicture}
    }\\
    \subfloat[\emph{Sound:} $N_1$ PMF]{        
        \begin{tikzpicture}
        \begin{axis}[
            pmfPlot,
            ylabel={$N_1$ PMF},
            xmin={0}, xmax={68}, xtick={1,10,20,30,40,50,60},
            ymin={0}, ymax={0.025}, ytick distance={0.005},
        ]
            \addplot [only marks,analytical,mark size=0.75pt,forget plot] table [x=n,y=PMF_Analytical] {data/N1_PMF_results_Sound.csv};

            \addplot [fill=simulated,draw=simulated] table [x=n,y=PMF_Simulation] {data/N1_PMF_results_Sound.csv};
        \end{axis}
        \end{tikzpicture}
    }
    \subfloat[\emph{Sound:} $\mathbb{E}\,\lbrack T_1\rbrack$ and $\mathrm{Var}(T_{1})$]{
        \begin{tikzpicture}
        \pgfplotsset{set layers}
        \begin{axis}[
            meanSubplot,
            ymin={250}, ymax={350}, ytick distance={25},
        ]
            \addplot [fill=analytical,draw=analytical!400,nodes near coords*={$\approx$\pgfmathprintnumber{\pgfplotspointmeta}}] coordinates {
                (Mean,311.8049291) (Variance,-999)
            };
            \addplot [fill=simulated,draw=simulated!300] coordinates {
                (Mean,311.84162) (Variance,-999)
            };
        \end{axis}
        \begin{axis}[
            varianceSubplot,
            ymin={5000}, ymax={25000}, ytick distance={5000},
        ]
            \addplot [fill=analytical,draw=analytical!400,nodes near coords*={$\approx$\pgfmathprintnumber{\pgfplotspointmeta}}] coordinates {
                (Mean,-999) (Variance,20041.563)
            };
            \addplot [fill=simulated,draw=simulated!300] coordinates {
                (Mean,-999) (Variance,20047.19574)
            };
        \end{axis}
        \end{tikzpicture}
    }
    \caption{{Comparison of the \gls{pmf} of $N_1$ and  moments of $T_1$ obtained analytically and through simulations. $D=100$~m; top row: $v=3\cdot10^{8}$~m/s, $C_1\in[10,500]$~ms; bottom row: $v=300$~m/s, $C_1\in[0,10]$~ms.}}\vspace{-0.5cm}
    \label{fig:t1_pmf_mean_grid}
\end{figure}
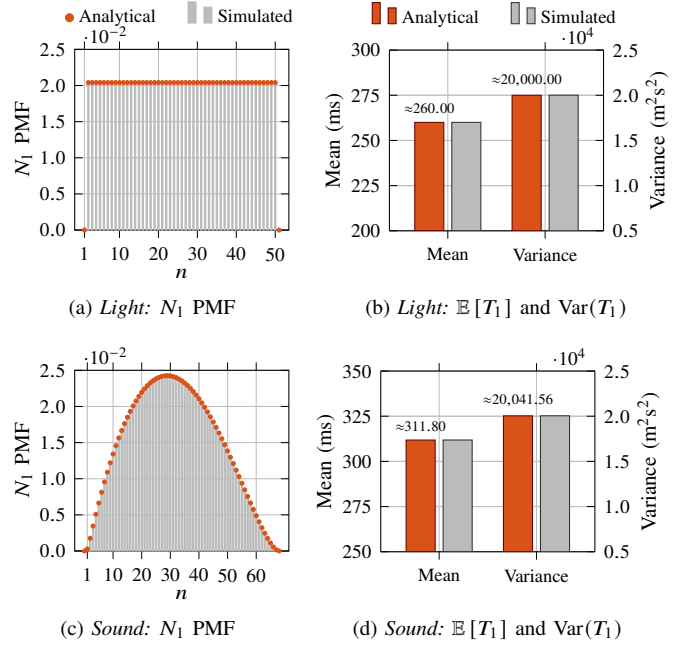

To address Problem~\eqref{eq:opt-twi-drop}, we first derive the \gls{pmf} of each delay component that constitutes $T_{\pi_k}$. The discreteness arises from frame quantization and retry mechanisms, which introduce ceiling operations. Crucially, these ceilings preclude exchanging expectation with the transformation, so moments cannot be obtained by linear averaging of the underlying continuous delays. Moreover, naive continuous approximations of the ceilinged sums are not straightforward and tend to be excessively permissive for reliability design. We therefore work directly with the induced discrete \glspl{pmf}, from which moments follow via finite sums.
\paragraph{Event-to-Sensor} We derive the \gls{pmf}, mean and variance of the first delay component in \( T_{\pi_k} \), denoted as:
\begin{equation}
T_1 = T_f \left\lceil \frac{T_{\text{ES}}^{\text{prop}} + C_1}{T_f} \right\rceil,
\end{equation}
where \( T_{\text{ES}}^{\text{prop}} \in [0, 2D/v] \) is the event-to-sensor propagation delay, and \( C_1 \sim \mathcal{U}[C_{\min}, C_{\max}] \) is the computation delay; see equations~\eqref{eq:ES_Prop_time} and \eqref{eq:Comp_time}. Next, we define the auxiliary sum:
\begin{equation}
Z := T_{\text{ES}}^{\text{prop}} + C_1,
\end{equation}
which lies in the bounded interval \( [C_{\min}, C_{\max} + 2D/v] \), as delineated in Appendix \ref{BDR_FS}. The \gls{pdf} of \( Z \), denoted \( f_Z(z) \), is given by the convolution of two independent random variables:
\begin{equation}
f_Z(z) = \int_{C_{\min}}^{C_{\max}} f_{T_{\text{ES}}^{\text{prop}}}(z - c) \cdot f_{C_1}(c) \, dc.
\end{equation}
To express the \gls{pmf} and expectation of \( T_1 \), we introduce the discrete random variable:
\begin{equation}
N_{1}:= \left\lceil \frac{Z}{T_f} \right\rceil \in \mathbb{N},
\end{equation}
where the former is given by:
\begin{equation}
\mathbb{P}(N_{1} = n) = \mathbb{P}\left( \frac{Z}{T_f} \in (n - 1, n] \right) 
= \int_{(n-1)T_f}^{nT_f} f_Z(z) \, dz,
\end{equation}
and the latter, \( \mathbb{E}[T_1] \) is subsequently computed as:
\begin{equation}
\mathbb{E}[T_1] = T_f \mathbb{E}[N_{1}] 
= T_f \sum_{n = n_{\min}}^{n_{\max}} n \mathbb{P}(N_{1} = n),
\end{equation}
with $n_{\max}$ and $n_{\min}$ defined in Appendix \ref{BDR_FS}. This derivation is numerically tractable by virtue of: (i) the support of \( Z \) is bounded, implying a finite number of summation terms; (ii) the ceiling operation is cleanly absorbed into discrete bins; and (iii) the integrals are over piecewise segments of \( f_Z \), which can be computed numerically. By extension, the variance of $T_1$ can be calculated as $\mathrm{Var}(T_{1})= T_{f}^{2}\,(\mathbb{E}[N_{1}^{2}]-\mathbb{E}[N_{1}]^2)$. We validate the derivation of the \gls{pmf}, the expectation, and the variance in Fig.~\ref{fig:t1_pmf_mean_grid}. It can be observed that with light as the physical signal, computation delay is dominant in $T_{1}$, whereas with sound, it is the propagation delay.

\begin{figure}[t]
    \centering


    \subfloat[$N_2$ PMF]{        
        \begin{tikzpicture}
        \begin{axis}[
            pmfPlot,
            ylabel={$N_2$ PMF},
            xmin={1}, xmax={75}, xtick={1,15,30,45,60,75},
            ymin={0}, ymax={0.025}, ytick distance={0.005},
            legend columns=2, legend style={at={(0.5,1.1)}, anchor=south, /tikz/every even column/.append style={column sep=1em}},
        ]
            \addplot [only marks,analytical,mark size=0.75pt,forget plot] table [x=n,y=PMF_Analytical] {data/N2_PMF_results_High_Comp.csv};
            \addlegendimage{line legend,only marks,mark size=1.5pt,color=analytical}
            \addlegendentry{Analytical}

            \addplot [fill=simulated,draw=simulated] table [x=n,y=PMF_Simulation] {data/N2_PMF_results_High_Comp.csv};
            \addlegendentry{Simulated}
        \end{axis}
        \end{tikzpicture}
    }
    \subfloat[$\mathbb{E}\,\lbrack T_2\rbrack$ and $\mathrm{Var}(T_{2})$]{
        \begin{tikzpicture}
        \pgfplotsset{set layers}
        \begin{axis}[
            meanSubplot,
            ymin={200}, ymax={300}, ytick distance={25},
            legend columns=2, legend style={at={(0.5,1.1)}, anchor=south, /tikz/every even column/.append style={column sep=1em}},
        ]
            \addplot [fill=analytical,draw=analytical!400,nodes near coords*={$\approx$\pgfmathprintnumber{\pgfplotspointmeta}}] coordinates {
                (Mean,274.2857143) (Variance,-999)
            };
            \addlegendentry{Analytical}
            \addplot [fill=simulated,draw=simulated!300] coordinates {
                (Mean,274.30873) (Variance,-999)
            };
            \addlegendentry{Simulated}
        \end{axis}
        \begin{axis}[
            varianceSubplot,
            ymin={5000}, ymax={25000}, ytick distance={5000},
        ]
            \addplot [fill=analytical,draw=analytical!400,nodes near coords*={$\approx$\pgfmathprintnumber{\pgfplotspointmeta}}] coordinates {
                (Mean,-999) (Variance,20061.22449)
            };
            \addplot [fill=simulated,draw=simulated!300] coordinates {
                (Mean,-999) (Variance,20057.45005)
            };
        \end{axis}
        \end{tikzpicture}
    }\\
    \subfloat[$N_2$ PMF]{        
        \begin{tikzpicture}
        \begin{axis}[
            pmfPlot,
            ylabel={$N_2$ PMF},
            xmin={0}, xmax={26}, xtick={1,10,20,30,40,50},
            ymin={0}, ymax={0.4}, ytick distance={0.1},
        ]
            \addplot [only marks,analytical,mark size=0.75pt,forget plot] table [x=n,y=PMF_Analytical] {data/N2_PMF_results_Low_Comp.csv};

            \addplot [fill=simulated,draw=simulated] table [x=n,y=PMF_Simulation] {data/N2_PMF_results_Low_Comp.csv};
        \end{axis}
        \end{tikzpicture}
    }
    \subfloat[$\mathbb{E}\,\lbrack T_2\rbrack$ and $\mathrm{Var}(T_{2})$]{
        \begin{tikzpicture}
        \pgfplotsset{set layers}
        \begin{axis}[
            meanSubplot,
            xshift=7pt,
            ymin={38}, ymax={46}, ytick distance={2},
        ]
            \addplot [fill=analytical,draw=analytical!400,nodes near coords*={$\approx$\pgfmathprintnumber{\pgfplotspointmeta}}] coordinates {
                (Mean,43.29980212) (Variance,-999)
            };
            \addplot [fill=simulated,draw=simulated!300] coordinates {
                (Mean,43.301415) (Variance,-999)
            };
        \end{axis}
        \begin{axis}[
            varianceSubplot,
            xshift=7pt,
            yticklabel style={/pgf/number format/zerofill=false},
            ymin={756}, ymax={772}, ytick distance={4},
        ]
            \addplot [fill=analytical,draw=analytical!400,nodes near coords*={$\approx$\pgfmathprintnumber{\pgfplotspointmeta}}] coordinates {
                (Mean,-999) (Variance,769.3938504)
            };
            \addplot [fill=simulated,draw=simulated!300] coordinates {
                (Mean,-999) (Variance,768.028309)
            };
        \end{axis}
        \end{tikzpicture}
    }
    \caption{{Comparison of the \gls{pmf} of $N_2$ and moments of $T_2$ obtained analytically and through simulations. $D=100$~m, $\rho_{k,i}=0.7$, $A_{\max}=25$; top row: $C_{k,i}\in[10,500]$ ms; 
    bottom row: $C_{k,i}\in[0,10]$ ms.}}\vspace{-0.5cm}
    \label{fig:t2_pmf_mean_grid}
\end{figure}
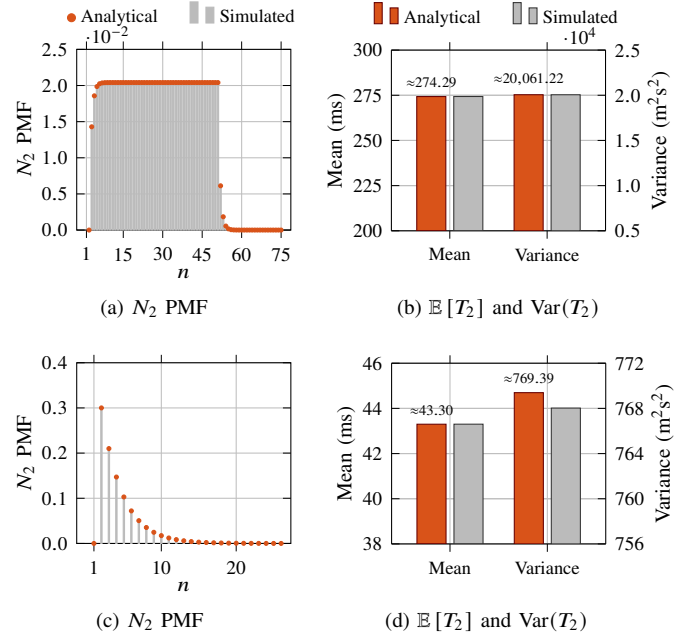

\paragraph{Sensor-to-Sensor hop} We now derive the \gls{pmf}, mean and variance of the second delay component, denoted \( T_2 \), which captures the sensor-to-sensor hop delay at node \( \pi_{k,i} \). This delay is modeled as:
\begin{equation}
T_2 = T_f \left\lceil a_{k,i} + \frac{C_{k,i}}{T_f} \right\rceil,
\end{equation}
where \( a_{k,i} \in \{1, 2, \dots, A_{\max}\} \) is a discrete random variable representing the number of communication frames required for successful hop transmission, with \gls{pmf} given by equation~\eqref{eq:T_hop_aki}. The variable \( C_{k,i} \sim \mathcal{U}[C_{\min}, C_{\max}] \) is the computation delay at node \( \pi_{k,i} \), and is independent of \( a_{k,i} \). We define the discrete random variable:
\begin{equation}
N_{2} := \left\lceil a_{k,i} + \frac{C_{k,i}}{T_f} \right\rceil \in \mathbb{N},
\end{equation}
whose \gls{pmf}, conditioned on each \( a \), $p_{a,n}=\mathbb{P}(N_{2} = n \mid a_{k,i} = a)$, is given by:
\begin{equation}
p_{a,n} = 
\frac{\min\{C_{\max}, T_f(n - a)\} - \max\{C_{\min}, T_f(n - 1 - a)\}}{C_{\max} - C_{\min}}.
\end{equation}
Then, the unconditional \gls{pmf} of \( N_{2} \) is:
\begin{equation}
\mathbb{P}(N_{2}=n) = \sum_{a=1}^{A_{\max}} \mathbb{P}(a_{k,i} = a) \cdot \mathbb{P}(N_{2} = n \mid a).
\end{equation}
Consequently, the expected value of \( T_2 \) is:
\begin{equation}
\mathbb{E}[T_2] = T_f \sum_{n} n \,\mathbb{P}(N_{2} = n),
\end{equation}
or, equivalently,
\begin{equation}
\mathbb{E}[T_2] =
T_f \sum_{a=1}^{A_{\max}}
\frac{\rho_{k,i}^{a-1}(1-\rho_{k,i})}{1 - \rho_{k,i}^{A_{\max}}}
\sum_{n=\lceil a+{C_{\min}/T_f} \rceil}^{\lceil a+{C_{\max}/T_f} \rceil}
n p_{a,n},
\end{equation}
where \( p_{a,n} = \mathbb{P}(N = n \mid a) \). Similar to $T_{1}$, the variance of $T_{2}$ is calculated in a straightforward way as $\mathrm{Var}(T_{2})= T_{f}^{2}\,(\mathbb{E}[N_{2}^{2}]-\mathbb{E}[N_{2}]^2)$. The full derivation is provided in Appendix~\ref{STS_DM}, and the analytical expression of the \gls{pmf}, mean, and variance is verified numerically in Fig.~\ref{fig:t2_pmf_mean_grid}.
Here, $N_{2}$ counts how many frame “steps” you need once you add a random computation delay to a random number of whole frames. For a fixed $a$, the uniform computation delay can push the sum either just over the next frame boundary or several boundaries ahead, so the chance of landing in each step $n$ is simply how much of $C_{k,i}$ falls into that step. Then we average these step–wise chances over all possible $a$’s: if the geometric distribution favors larger $a$, the whole \gls{pmf} shifts to the right. A wider computation range $[C_{\min},C_{\max}]$ spreads probability across more steps (flatter, wider \gls{pmf}), while a larger frame time $T_f$ uses coarser steps so mass concentrates into fewer bins. 
Because both $a_{k,i}$ and $C_{k,i}$ are bounded, the \gls{pmf} has a finite range and naturally tapers at the edges.
\paragraph{Sensor-to-BS} The third component, i.e., sensor-to-\gls{bs} delay comprises of two truncated geometric random variables, $T^{\text{SR}}_{k,i}$ and $T^{\text{PT}}_{k,i}$. Their respective \gls{pmf}s are delineated in Section \ref{Sys_Mod}, and means can be calculated in a straightforward way as:
\begin{equation}
\begin{split}
\mathbb{E}[T^{\text{SR}}_{k,i}] &= T_f \frac{1 - (M_{\max}+1)\,\zeta_{k,i}^{M_{\max}} + M_{\max} \, \zeta_{k,i}^{M_{\max}+1}}{(1 - \zeta_{k,i}) (1 - \zeta_{k,i}^{M_{\max}})}, \\
\mathbb{E}[T^{\text{PT}}_{k,i}] &= T_f \frac{1 - (N_{\max}+1)\,\epsilon_{k,i}^{N_{\max}} + N_{\max} \, \epsilon_{k,i}^{N_{\max}+1}}{(1 - \epsilon_{k,i}) (1 - \epsilon_{k,i}^{N_{\max}})}. \\
\end{split}
\end{equation}
Similarly, variance of the two is obtained in Appendix \ref{Var_TSR_TPT}. 
\par After deriving the \gls{pmf} of the individual components, the \gls{pmf} of the total delay is obtained as the {discrete convolution} of these component \glspl{pmf}. We note, however, that the tractability of the resulting convolutions, and whether they admit closed forms or resemble a known family (at least in moments), depends on the parameter regime (e.g., frame time, computation ranges, geometric retry parameters and truncation), as well as independence assumptions between components. Moreover, in regimes with wide supports or pronounced skew, exact analytical simplifications may be limited. Consequently, we characterize satisfaction probabilities and guarantees via known probabilistic bounds. For tractability and ease of exposition, we work with the first two moments (mean and variance); however, since the component \glspl{pmf} are available, higher-order moments can be incorporated to tighten these bounds when warranted~\cite{Tight_Cantelli}.
\par To this end, per-path timeliness requires a computable lower bound on
\(\Pr\!\big(T_{\pi_k}\le \text{\gls{twi}}\,\big|\,\text{no drop}\big)\)
that is (i) tight, (ii) distribution-agnostic, and (iii) amenable to convex optimization. We therefore use the one-sided Chebyshev (Cantelli) inequality, which meets these criteria. For any random variable \(S\) with mean \(\mu\) and variance \(\sigma^2\), and any \(w>0\):
\begin{equation}
\Pr\!\left(S \le \mu + w\right) \;\ge\; \frac{w^2}{\sigma^2 + w^2}.
\label{eq:cantelli}
\end{equation}
We adopt the Cantelli bound because, among distribution-free inequalities that use only the first two moments, it is sharp, so no uniformly tighter one-sided bound exists without additional structure~\cite{Tight_Cantelli}. In contrast, Markov bounds depend only on the mean and are typically too loose for latency design, while Chernoff bounds can be tighter but require additional  structural information, and can introduce a non-convex optimization over the tilt parameter. By using only \((\mu,\sigma^2)\), which we obtain in closed form for each delay component, Cantelli yields computable, convex-friendly constraints.
\par Let $S_k \equiv \{T_{\pi_k}\,|\,\text{no drop}\}$ denote the end-to-end delay of path $\pi_k$ conditioned on successful delivery, with mean $\mu_k$ and standard deviation $\sigma_k$. Moreover, let $p_k \triangleq P^{\text{drop}}_{\pi_k}$ denote the path-level drop probability; as calculated in equation \eqref{eq:drop-path}. Then, for any target level $\alpha_k\in(0,1)$, we require
$\Pr(S_k \le \text{TWI}) \ge \alpha_k$. Applying Cantelli’s inequality in \eqref{eq:cantelli} with $w=\text{TWI}-\mu_k>0$ yields the deterministic margin:
\begin{equation}
\text{TWI} \ge \mu_k + \sigma_k\sqrt{\frac{\alpha_k}{\,1-\alpha_k\,}},
\label{eq:cantelli_margin}
\end{equation}
where $\alpha_k=(1-\varepsilon_k)/(1-p_k)$. The margin factor $\theta(\alpha)\!\triangleq\!\sqrt{\alpha/(1-\alpha)}$ is strictly increasing and convex on $(0,1)$; hence tighter reliability (larger $\alpha$) implies a larger deterministic safety margin added to $\mu_k$. Next, substituting \eqref{eq:cantelli_margin} in Problem \eqref{eq:opt-twi-drop}, we obtain the following design problem:
\begin{subequations}\label{eq:opt_alpha_native}
\begin{align}
\underset{\text{TWI},\,\{\varepsilon_k\}}{\text{minimise}} \quad& \text{TWI} \\
\text{s.t.}\quad
& \text{TWI} \ge \mu_k + \sigma_k, \theta(\alpha_k),\quad\forall k,  
\label{eq:cantelli_native}\\
& \sum_{k=1}^K \varepsilon_k \le \varepsilon,\label{eq:budget_native_sum}\\
&\varepsilon_k \ge\ p_k,\quad\forall k.
\label{eq:budget_native}
\end{align}
\end{subequations}
\begin{algorithm}[b]
\caption{Bisection for Minimum \gls{twi}}
\label{alg:twi_bisection_simple}
\footnotesize
\textbf{Inputs:} $\{\mu_k,\sigma_k^2,p_k\mid \forall k\}$, $\varepsilon$, $\iota$.\\
\textbf{Feasibility check:} If $\sum_{k=1}^K p_k > \varepsilon$, declare infeasible and stop.\\
\textbf{Initialize bracket:} $[\text{TWI}_{\min},\text{TWI}_{\max}]$\\
\textbf{Repeat until} $\text{TWI}_{\max}-\text{TWI}_{\min} \le \iota$:
  \begin{enumerate}
    \item Set $\text{TWI} \leftarrow (\text{TWI}_{\min}+\text{TWI}_{\max})/2$.
    \item For each $k=\{1,\dots,K\}$, compute
          \begin{equation*}
            y_k \leftarrow \max\{\text{TWI}-\mu_k,\,0\},\
          \bar{\alpha}_k \leftarrow 
          {y_k^2}/({y_k^2+\sigma_k^2}), 
          \end{equation*}
          then clip
          \[
          \alpha_k^\star \leftarrow \min\{\,1,\ \max\{\alpha_{\min},\,\bar{\alpha}_k\}\,\}.
          \]
    \item Evaluate: $\Phi(\text{TWI}) \leftarrow \sum_{k=1}^K (1-p_k)\,\alpha_k^\star.$
    \item \textit{If} $\Phi(\text{TWI}) \ge K-\varepsilon$: \textbf{set} $\text{TWI}_{\max} \leftarrow \text{TWI}$ \\
          \textit{else}: \textbf{set} $\text{TWI}_{\min} \leftarrow \text{TWI}$.
  \end{enumerate}
\textbf{Return:} $\text{TWI}^\star$,
        $\alpha_k^\star(\text{TWI}^\star)$,
        $\varepsilon_k^\star \leftarrow 1-(1-p_k)\alpha_k^\star$.
\end{algorithm}
Problem \eqref{eq:opt_alpha_native} can be solved via a quasi–convex reduction and a bisection search on \gls{twi}.
Let the Cantelli margin be $\theta(\alpha)=\sqrt{\alpha/(1-\alpha)}$. The objective is linear in \gls{twi}. For
$\alpha\in[1/4,1]$, $\theta(\alpha)$ is convex; hence the Cantelli constraints in \eqref{eq:cantelli_native} are convex in
$(\text{TWI},\alpha_k)$ on that domain. The drop–budget constraint \eqref{eq:budget_native} becomes linear after substituting
$\varepsilon_k = 1-(1-p_k)\alpha_k$, yielding:
\begin{equation}
\sum_{k=1}^{K} (1-p_k)\,\alpha_k \;\ge\; K-\varepsilon .
\label{eq:alpha_budget}
\end{equation}
Next, fix a candidate $\text{TWI}$ and define the non-negative margin
$y_k(\text{TWI})\triangleq\max\{\text{TWI}-\mu_k,\,0\}$. Then \eqref{eq:cantelli_native} is equivalent to the per–path {upper bound}
\begin{equation}
\alpha_k \;\le\; \bar\alpha_k(\text{TWI})\;\triangleq\;\frac{y_k(\text{TWI})^2}{\,y_k(\text{TWI})^2+\sigma_k^2\,},\, \forall k.
\label{eq:alpha_cap}
\end{equation}
Note that, we also have a reliability floor $\alpha_k\ge \alpha_{\min}=\tfrac14$. For fixed $\text{TWI}$, the choice that most helps the linear budget in equation \eqref{eq:alpha_budget} is therefore
\begin{equation}
\alpha_k^\star(\text{TWI})\;=\;\min\Big\{1,\ \max\big\{\alpha_{\min},\ \bar\alpha_k(\text{TWI})\big\}\Big\}.
\label{eq:alpha_star}
\end{equation}
Following this, we define the feasibility map
\begin{equation}
\Phi(\text{TWI})\;\triangleq\;\sum_{k=1}^{K} (1-p_k)\,\alpha_k^\star(\text{TWI}).
\label{eq:phi_def}
\end{equation}
Because each $\bar\alpha_k(\text{TWI})$ is nondecreasing in $\text{TWI}$, $\Phi(\text{TWI})$ is nondecreasing; moreover
$\lim_{\text{TWI}\to\infty}\Phi(\text{TWI})=\sum_k (1-p_k)=K-\sum_k p_k$. Hence, if the necessary condition
$\sum_k p_k\le \varepsilon$ holds, there exists a finite $\text{TWI}$ with $\Phi(\text{TWI})\ge K-\varepsilon$.
\begin{algorithm}[t]
\caption{Uniform Baseline for \gls{twi}}
\label{alg:twi_uniform}
\footnotesize
\textbf{Inputs:} $\{\mu_k,\sigma_k^2,p_k\mid \forall k\}$, total budget $\varepsilon$.\\
\textbf{Feasibility check:} If $\sum_{k=1}^K p_k > \varepsilon$, declare infeasible and stop.\\
\textbf{Find $\tau$:} Choose $\tau$ such that $\sum_{k=1}^K \max\{p_k,\tau\}=\varepsilon$
using bisection on $\tau$.\\
\textbf{Set:} $\varepsilon_k \leftarrow \max\{p_k,\tau\}$, $\forall k$.\\
\textbf{Compute:} $\alpha_k \leftarrow {(1-\varepsilon_k)}/{(1-p_k)}$, for all $k$.\\
\textbf{Compute:} $ \sigma_k \theta_k $, $\forall k$.\\
\textbf{Baseline window:} $\text{TWI}_{\text{uni}} \leftarrow \max_k \{\mu_k + \sigma_k \theta_k \}$.\\
\textbf{Return:} $\text{TWI}_{\text{uni}}$, $\{\alpha_k\}$, $\{\varepsilon_k\}$.
\end{algorithm}
\par Finally, we solve \eqref{eq:opt_alpha_native} by bisection on $\text{TWI}$ over an interval
$[\,\text{TWI}_{\min},\,\text{TWI}_{\max}\,]$, where $\text{TWI}_{\min}=\max_k \mu_k$ and $\text{TWI}_{\max}$ is any
sufficiently large value for which $\Phi(\text{TWI}_{\max})\ge K-\varepsilon$ holds. At each iteration we test the midpoint $\text{TWI}$:
compute $\alpha_k^\star(\text{TWI})$ via \eqref{eq:alpha_star}, evaluate $\Phi(\text{TWI})$ in \eqref{eq:phi_def}, and keep the lower
(respectively upper) subinterval if $\Phi(\text{TWI})< K-\varepsilon$ (respectively $\Phi(\text{TWI})\ge K-\varepsilon$). Since the feasible set in
$\text{TWI}$ is an interval by monotonicity, bisection returns the {global} optimum $\text{TWI}^\star$. The associated reliability
allocation is $\alpha_k^\star=\alpha_k^\star(\text{TWI}^\star)$, and the per–path violation levels recover as
$\varepsilon_k^\star = 1-(1-p_k)\,\alpha_k^\star$. The algorithm runs to a tolerance level $\iota$. Algorithm \ref{alg:twi_bisection_simple} delineates the bisection optimisation algorithm to obtain minimum \gls{twi}.
\begin{table}[b]
\centering
\caption{Simulation Parameters}
\label{tab:params}
\begin{tabular}{ll}
\toprule
\textbf{Parameter} & \textbf{Value / Range} \\
\midrule
$T_f$                                 & $10$ ms \\
$D$                                   & $100$ m\\
$\varepsilon$                          & $0.1$ \\
$T1_{\text{light}}$                    & $v = 3\times10^{8}/10^{3},\;\; C_{1,\min}=10,\; C_{1,\max}=200$ \\
$T1_{\text{sound}}$                    & $v = 300/10^{3},\;\; C_{1,\min}=0,\; C_{1,\max}=10$ \\
$T2$                                   & $C_{k,i} \in [0,10],\;\; \rho_{k,i} \in [0.05,0.15],\;\; A_{\max} \in \{2,\dots,5\}$ \\
$T^{\text{SR}}$                       & $\zeta_{k,i} \in [0.05,0.10],\;\; M_{\max} \in \{2,\dots,5\}$ \\
$T^{\text{PT}}$                       & $\epsilon_{k,i} \in [0.01,0.10],\;\; N_{\max} \in \{3,\dots,7\}$ \\
\bottomrule
\end{tabular}
\end{table}
\par As a conservative baseline, we allocate the violation budget by raising all per-path violations to a common level above their intrinsic drops. Concretely, choose a threshold \(\tau\) and set \(\varepsilon_k=\max\{p_k,\tau\}\) \(\forall k\), with \(\sum_{k=1}^K \varepsilon_k=\varepsilon\). This yields a water-filling allocation over the floors \(\{p_k\}\). Define \(S(\tau)=\sum_{k=1}^K \max\{p_k,\tau\}\). On \(\tau\in[\min_k p_k,1)\), \(S(\tau)\) is piecewise-linear and strictly increasing, with \(S(\min_k p_k)=\sum_k p_k\) and \(S(1)=K\). Hence, for any feasible budget \(\sum_k p_k \le \varepsilon < K\), there exists a unique \(\tau\in[\min_k p_k,1)\) solving \(S(\tau)=\varepsilon\) (e.g., by bisection). The induced conditional reliabilities are \(\alpha_k = (1-\varepsilon_k)/(1-p_k) \in (0,1)\), giving a Cantelli margin \(\theta(\alpha_k)=\sqrt{\alpha_k/(1-\alpha_k)}\) on path \(k\). The baseline \gls{twi} is then computed as:
\begin{equation}
\text{TWI}_{\text{uni}}
\;=\;
\max_{k=1,\dots,K}
\Big\{\,
\mu_k + \sigma_k \sqrt{\tfrac{\alpha_k}{\,1-\alpha_k\,}}
\Big\}.
\end{equation}
This construction yields a feasible pair $(\text{TWI}_{\text{uni}},\{\varepsilon_k\})$ whenever $\sum_k p_k\le \varepsilon$. Algorithm \ref{alg:twi_uniform} delineates the uniform allocation algorithm for $\varepsilon_k$ and, in turn, associated \gls{twi}.
\section{Numerical Results}\label{Num_Results}
In this section, we evaluate timeliness under the proposed design using Monte Carlo simulations. Unless stated otherwise, parameters follow Table~\ref{tab:params}~\cite{Joao_SV}. Each experiment/run reports the achievable \gls{twi} ($\Delta$)in ms together with the allocation \(\{\varepsilon_k\}\) under a global budget \(\sum_k \varepsilon_k \le \varepsilon\). Because hop- and link-level parameters are randomly drawn in each run, we report averages over $1000$ independent realizations. For feasibility at the union-bound edge, we set \(\varepsilon \leftarrow \max\{\varepsilon,\sum_k p_k + 10^{-4}\}\) per realization. We compare two policies: (i) \textbf{Optimal}, obtained using Algorithm \ref{alg:twi_bisection_simple}, and (ii) \textbf{Uniform}, using Algorithm \ref{alg:twi_uniform}.  We first present results for specific fixed path compositions, and then turn to randomized path topologies to assess robustness.

\subsection{Specific Scenarios}
We begin with three canonical two-path, $\mathcal{P}=\{\pi_1,\pi_2\}$, scenarios that capture the basic ways information about a physical event can reach the \gls{bs}. {Scenario-A} pairs a {direct} branch $\pi_1=\{BS\}\Leftrightarrow\{\text{Event}\rightarrow BS\}$ with a {relayed} branch $\pi_2=\{s_{2,1}\rightarrow BS\}\Leftrightarrow\{\text{Event}\rightarrow \text{Sensor}\rightarrow BS\}$, both using light propagation. {Scenario-B} mirrors this in sound. {Scenario-C} removes the direct route and considers two relayed branches in parallel, one with light and one with sound: $\pi_1=\{s_{1,1}\rightarrow BS\}\Leftrightarrow\{\text{Event}\rightarrow \text{Sensor (light)}\rightarrow BS\}$ and $\pi_2=\{s_{2,1}\rightarrow BS\}\Leftrightarrow\{\text{Event}\rightarrow \text{Sensor (sound)}\rightarrow BS\}$. These foundational scenarios allow us to cleanly dissect how the presence of a direct path, propagation modality (light versus sound in our case), and hop-induced computation/access variability shape the computed \gls{twi} (usefulness horizon) and the budget allocation \(\{\varepsilon_k\}\), thereby clarifying which factors drive timeliness, and how strongly. In the following, each figure corresponds to one of the three specific-path scenarios and contains two plots: on the left, the achievable \gls{twi}, with bars showing means and caps indicating one standard deviation; on the {right}, the average per-path budget allocations \(\{\epsilon_k\}\) under the same two policies, i.e., Optimal and Uniform.
\par
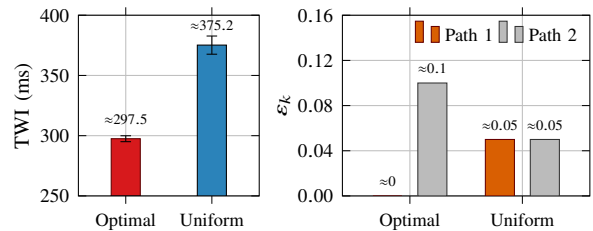
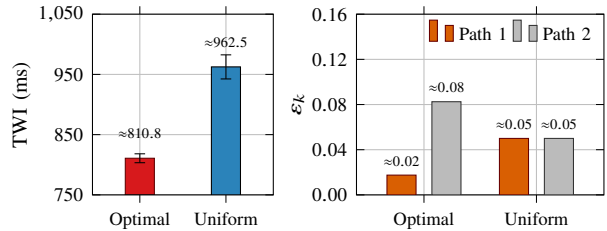
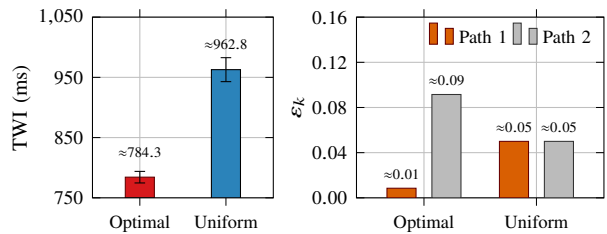
\begin{figure}[t]
    \centering
    \subfloat[Scenario-A]{
        \begin{tikzpicture}
        \begin{axis}[
            twiBarPlot,
            ymin={250}, ymax={400}, ytick distance={50},
            nodes near coords style={yshift=2pt},
        ]
            \addplot [
                fill=twi_optimal,draw=twi_optimal!400,
                error bars/.cd,y dir=both,y explicit,error bar style={black},
            ] coordinates {
                (Optimal,297.5) +- (2.46,2.46)
                (Uniform,-999) +- (0,0)
            };
            \addplot [
                fill=twi_uniform,draw=twi_uniform!400,
                error bars/.cd,y dir=both,y explicit,error bar style={black},
            ] coordinates {
                (Uniform,375.2) +- (7.55,7.55)
                (Optimal,-999) +- (0,0)
            };
        \end{axis}
        \end{tikzpicture}
        \begin{tikzpicture}
        \begin{axis}[
            reliabilityBarPlot,
            ymin={0}, ymax={0.16}, ytick distance={0.04}
        ]
            \addplot [fill=path1,draw=path1!400] coordinates {(Optimal,0) (Uniform,0.05)};
            \addlegendentry{Path 1}
            
            \addplot [fill=path2,draw=path2!300] coordinates {(Optimal,0.1) (Uniform,0.05)};
            \addlegendentry{Path 2}
        \end{axis}
        \end{tikzpicture}
    }\\
    \subfloat[Scenario-B]{
        \begin{tikzpicture}
        \begin{axis}[
            twiBarPlot,
            ymin={750}, ymax={1050}, ytick={750,850,950,1050},
            nodes near coords style={yshift=4pt},
        ]
            \addplot [
                fill=twi_optimal,draw=twi_optimal!400,
                error bars/.cd,y dir=both,y explicit,error bar style={black}
            ] coordinates {
                (Optimal,810.8) +- (7.4,7.4)
                (Uniform,-999) +- (0,0)
            };
            \addplot [
                fill=twi_uniform,draw=twi_uniform!400,
                error bars/.cd,y dir=both,y explicit,error bar style={black}
            ] coordinates {
                (Uniform,962.5) +- (20,20)
                (Optimal,-999) +- (0,0)
            };
        \end{axis}
        \end{tikzpicture}
        \begin{tikzpicture}
        \begin{axis}[
            reliabilityBarPlot,
            ymin={0}, ymax={0.16}, ytick distance={0.04},
        ]
            \addplot [fill=path1,draw=path1!400] coordinates {(Optimal,0.0175) (Uniform,0.05)};
            \addlegendentry{Path 1}
            
            \addplot [fill=path2,draw=path2!300] coordinates {(Optimal,0.0825) (Uniform,0.05)};
            \addlegendentry{Path 2}
        \end{axis}
        \end{tikzpicture}
    }\\
    \subfloat[Scenario-C]{
        \begin{tikzpicture}
        \begin{axis}[
            twiBarPlot,
            ymin={750}, ymax={1050}, ytick={750,850,950,1050},
            nodes near coords style={yshift=4pt},
        ]
            \addplot [
                fill=twi_optimal,draw=twi_optimal!400,
                error bars/.cd,y dir=both,y explicit,error bar style={black},
            ] coordinates {
                (Optimal,784.3) +- (9.6,9.6)
                (Uniform,-999) +- (0,0)
            };
            \addplot [
                fill=twi_uniform,draw=twi_uniform!400,
                error bars/.cd,y dir=both,y explicit,error bar style={black},
            ] coordinates {
                (Uniform,962.8) +- (19.9,19.9)
                (Optimal,-999) +- (0,0)
            };
        \end{axis}
        \end{tikzpicture}
        \begin{tikzpicture}
        \begin{axis}[
            reliabilityBarPlot,
            ymin={0}, ymax={0.16}, ytick distance={0.04},
        ]
            \addplot [fill=path1,draw=path1!400] coordinates {(Optimal,0.0085) (Uniform,0.05)};
            \addlegendentry{Path 1}
            
            \addplot [fill=path2,draw=path2!300] coordinates {(Optimal,0.0915) (Uniform,0.05)};
            \addlegendentry{Path 2}
        \end{axis}
        \end{tikzpicture}
    }
    \caption{\gls{twi} and per-path violation budget computed by the optimal algorithm (Algorithm~\ref{alg:twi_bisection_simple}) and the uniform baseline (Algorithm~\ref{alg:twi_uniform}).}\vspace{-0.8cm}
    \label{fig:BS_Onehop}
\end{figure}

 In Fig.~\ref{fig:BS_Onehop}(a), the optimal \gls{twi} Algorithm~\ref{alg:twi_bisection_simple} achieves a markedly smaller \gls{twi} than uniform Algorithm~\ref{alg:twi_uniform}: \(297.5\) versus  \(375.2\)~ms, on average a \(20.7\%\) reduction, with lower variability (standard deviation \(2.52\) versus\ \(7.66\)~ms). The reason is visible in the budget splits. {Uniform} allocation fixes \(\epsilon_k=[0.05,\,0.05]\), forcing both branches to meet a similar conditional timeliness target, so the relayed branch, which carries most of the delay variance due to computation/access, must sustain a tight Cantelli margin, driving up the max that defines \gls{twi}. In contrast, {optimal} allocation concentrates essentially the entire budget on the relayed branch, \(\epsilon_k\approx[0,\,0.10]\) (direct path \(\approx 0\); relayed \(\approx 0.10\)), which relaxes its conditional timeliness requirement and sharply reduces its required margin. The direct light path, being low-variance and short, remains timely even under a stricter target and therefore ceases to be the bottleneck. As a result, the worst-path margin drops, and so does the overall \gls{twi}.
\par Next, Fig.~\ref{fig:BS_Onehop}(b) illustrates the \gls{twi} for Scenario-B. Relative to Scenario-A, both policies yield substantially larger \glspl{twi} in Scenario-B: sound propagation and the associated timing variability inflate the worst-path Cantelli margin. Still, Algorithm~\ref{alg:twi_bisection_simple} outperforms Algorithm~\ref{alg:twi_uniform} by a significant margin with a \(\!15.9\%\) reduction, with lower variability. Again, the reliability splits explain the gap: {Uniform} fixes \(\epsilon_k=[0.05,\,0.05]\), forcing both the direct and relayed sound branches to meet similar conditional timeliness, which keeps the tighter (noisier) margin high. {Optimal} shifts most of the lateness budget to the relayed branch while assigning a smaller, but nonzero, share to the direct branch, \(\epsilon_k\approx[0.0175,\,0.0825]\). Unlike Scenario~A (where the direct light branch could absorb almost no budget), the direct \emph{sound} branch requires a bit of slack to prevent it from becoming the bottleneck; the remaining slack on the relayed sound branch lowers the overall worst-path margin, hence the reduced \gls{twi}.
\par Finally, in Fig.~\ref{fig:BS_Onehop}(c), with both branches relayed, \glspl{twi} are high overall because each path incurs sensor–side computation and \gls{sr}/\gls{pt} access variability, and the sound branch becomes the bottleneck. Even so, Algorithm~\ref{alg:twi_bisection_simple} reduces \gls{twi} to \(784.3\)~ms on average versus \(962.8\)~ms for Algorithm~\ref{alg:twi_uniform}. The reliability allocations follow the same total budget \(\varepsilon=0.1\) as in the other scenarios; {Uniform} keeps equal splits \([0.05,\,0.05]\), while {Optimal} concentrates the budget on the harder (sound) path \([\sim0.01,\,\sim 0.09]\), tightening the light branch and lowering the worst-path Cantelli margin that determines \gls{twi}.

\begin{figure*}[!t]
    \centering
    \subfloat[$H_{\max}=1$]{
        \begin{tikzpicture}
        \begin{axis}[
            twiLinePlot,
            xlabel={$K$},
            ylabel={TWI (ms)},
            xmin={2}, xmax={5}, xtick distance={1},
            ymin={0}, ymax={2000}, ytick distance={500},
            legend columns=2, legend style={at={(0.5,1.1)}, anchor=south, /tikz/every even column/.append style={column sep=1em}}
        ]
            \addplot [optimalLine] coordinates {
                (2,652.4) (3,794.6) (4,910.5) (5,1004.8)
                
            };
            \addlegendentry{Optimal}
            
            \addplot [uniformLine] coordinates {
                (2,761.7) (3,999.4) (4,1201.2) (5,1376)
            };
            \addlegendentry{Uniform}

            \addplot[name path=optimal_uppbound,draw=none] coordinates {
                (2,1013.2) (3,1175.1) (4,1309.3) (5,1393)
            };
            \addplot[name path=optimal_lowbound,draw=none,domain={1:12}] coordinates {
                (2,295.3) (3,370.3) (4,426.7) (5,474.5)
            };
            \addplot[optimalArea] fill between [of=optimal_uppbound and optimal_lowbound];

            \addplot[name path=uniform_uppbound,draw=none] coordinates {
                (2,1026.4) (3,1251.4) (4,1504.3) (5,1798.7)
            };
            \addplot[name path=uniform_lowbound,draw=none,domain={1:12}] coordinates {
                (2,369.8) (3,426.7) (4,474.4) (5,517.9)
            };
            \addplot[uniformArea] fill between [of=uniform_uppbound and uniform_lowbound];
        \end{axis}
        \end{tikzpicture}
    }
    \subfloat[$K=3$]{
        \begin{tikzpicture}
        \begin{axis}[
            twiLinePlot,
            xlabel={$H_{\max}$},
            ylabel={TWI (ms)},
            xmin={1}, xmax={4}, xtick distance={1},
            ymin={0}, ymax={2000}, ytick distance={500},
        ]
            \addplot [optimalLine] coordinates {
                (1,794.6) (2,814.3) (3,850.2) (4,877.8)
            };
            
            \addplot [uniformLine] coordinates {
                (1,999.4) (2,1035.5) (3,1312.3) (4,1713.4)
            };

            \addplot[name path=optimal_uppbound,draw=none] coordinates {
                (1,1175.1) (2,1357) (3,1665.2) (4,1641.1)
            };
            \addplot[name path=optimal_lowbound,draw=none,domain={1:12}] coordinates {
                (1,370.3) (2,370.1) (3,370.5) (4,370.6)
            };
            \addplot[optimalArea] fill between [of=optimal_uppbound and optimal_lowbound];

            \addplot[name path=uniform_uppbound,draw=none] coordinates {
                (1,1251.4) (2,3539.6) (3,45189.6) (4,45219.4)
            };
            \addplot[name path=uniform_lowbound,draw=none,domain={1:12}] coordinates {
                (1,426.7) (2,426.7) (3,426.9) (4,427.7)
            };
            \addplot[uniformArea] fill between [of=uniform_uppbound and uniform_lowbound];
        \end{axis}
        \end{tikzpicture}
    }
    \subfloat[$D=100$~m]{
        \begin{tikzpicture}
        \begin{axis}[
            twiLinePlot,
            xlabel={$T_f$~(ms)},
            ylabel={TWI (ms)},
            xmin={4}, xmax={22}, xtick={4,10,16,22},
            ymin={0}, ymax={2000}, ytick distance={500},
        ]
            \addplot [optimalLine] coordinates {
                (4, 808.7) (10,820.5) (16,840.4) (22,852.7)
            };
            
            \addplot [uniformLine] coordinates {
                (4,1028.7) (10,1041.9) (16,1056.3) (22,1063.8)
            };

            \addplot[name path=optimal_uppbound,draw=none] coordinates {
                (4,1270.2) (10,1208.5) (16,1296.1) (22,1294)
            };
            \addplot[name path=optimal_lowbound,draw=none,domain={1:12}] coordinates {
                (4,354.6) (10,370.5) (16,388.2) (22,405.1)
            };
            \addplot[optimalArea] fill between [of=optimal_uppbound and optimal_lowbound];

            \addplot[name path=uniform_uppbound,draw=none] coordinates {
                (4,3605.1) (10,2631.3) (16,2643.6) (22,1912.1)
            };
            \addplot[name path=uniform_lowbound,draw=none,domain={1:12}] coordinates {
                (4,411) (10,426.9) (16,445.5) (22,462.6)
            };
            \addplot[uniformArea] fill between [of=uniform_uppbound and uniform_lowbound];
        \end{axis}
        \end{tikzpicture}
    }
    \subfloat[$T_f=10$~ms]{
        \begin{tikzpicture}
        \begin{axis}[
            twiLinePlot,
            xlabel={$D$~(m)},
            ylabel={TWI (ms)},
            xmin={50}, xmax={200}, xtick={50,100,150,200},
            ymin={0}, ymax={4000}, ytick distance={1000},
        ]
            \addplot [optimalLine] coordinates {
                (50,486.9) (100,815.9) (150,1158.6) (200,1519)
            };
            
            \addplot [uniformLine] coordinates {
                (50,576.4) (100, 1034.4) (150,1481.4) (200, 1981.4)
            };

            \addplot[name path=optimal_uppbound,draw=none] coordinates {
                (50,702.9) (100,1281) (150,1839.3) (200,2460.4)
            };
            \addplot[name path=optimal_lowbound,draw=none,domain={1:12}] coordinates {
                (50,370.1) (100,370.6) (150,370.4) (200,370.1)
            };
            \addplot[optimalArea] fill between [of=optimal_uppbound and optimal_lowbound];

            \addplot[name path=uniform_uppbound,draw=none] coordinates {
                (50,1800.7) (100,2286.2) (150,2745.5) (200,4202.9)
            };
            \addplot[name path=uniform_lowbound,draw=none,domain={1:12}] coordinates {
             (50,426.5) (100,427.4) (150,426.8) (200,426.6)
            };
            \addplot[uniformArea] fill between [of=uniform_uppbound and uniform_lowbound];
        \end{axis}
        \end{tikzpicture}
    }
    \caption{\Gls{twi} obtained by varying (a)~$K$, (b)~$H_{\max}$, (c)~$T_f$, and (d) cell size~$D$.}\vspace{-0.1cm}
    \label{fig:Varying_Params_TWI}
\end{figure*}
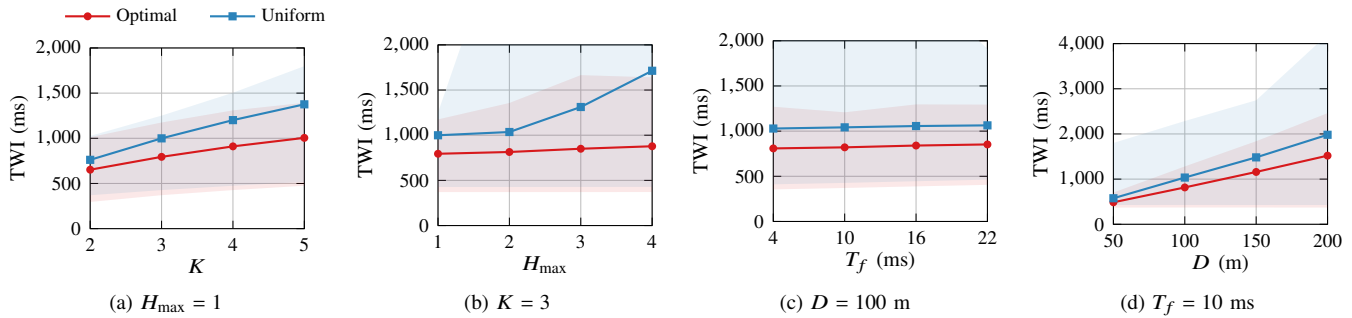
\vspace{-0.1cm}
\subsection{General Scenarios}
In this subsection, we study and analyse how timeliness varies with key system parameters by running Monte Carlo experiments in which, for each parameter setting, we generate $1000$ independent network realizations. In each realization, processing/link parameters are drawn at random from the ranges in Table~\ref{tab:params}. Moreover,  we cap the number of sensors per path by \(H_{\max}\).
For each path \(k\), we draw the sensor count \(h_k\) uniformly from the discrete set
\(\{0,1,\ldots,H_{\max}\}\). When \(h_k=0\), the path reduces to direct \gls{bs} sensing,
\(\pi_k=\{\text{\gls{bs}}\}\); otherwise \(\pi_k=\{s_{k,1}\rightarrow\cdots\rightarrow s_{k,h_k}\rightarrow \text{\gls{bs}}\}\).
Thus, a range \([0,3]\) allows paths with 0 up to 3 sensors. For practical purposes, across multiple paths, we
admit at most one direct event–to–\gls{bs} path (\(h_k=0\)). Subsequently, for a fixed global budget $\varepsilon$, we compute the \gls{twi}. For the general scenario, reported curves show the mean \gls{twi}, with a shaded band indicating the empirical min–max spread over the $1000$ draws.
\par In Fig.~\ref{fig:Varying_Params_TWI}(a), as we increase the number of candidate paths from $K=2$ to $K=5$ while keeping the range of $h_k$ fixed to $[0,{H}_{\max}{=}1]$, both policies exhibit larger \gls{twi}: the mean rises from $652.4$ to $1004.2$\,ms for {Optimal} and from $761.7$ to $1376$\,ms for {Uniform}. The \gls{twi} gap between the two widens with $K$, approximately $14\%$ at $K{=}2$ to $27\%$ at $K{=}5$, consistent with Optimal algorithm reallocating the budget $\{\varepsilon_k\}$ toward the hardest path and avoiding $\alpha_k\!\to\!1$ at the bottleneck. The empirical {min--max} envelope is also much broader for {Uniform}, and expands as $K$ increases, indicating heavier tails and occasional outliers, while {Optimal} remains comparatively tight. Next, 
in Fig.~\ref{fig:Varying_Params_TWI}(b), fixing $K{=}3$ and expanding the sensor from $[0,H_{\max}=1]$ to $[0,H_{\max}=4]$ leaves {Optimal} algorithm fairly stable with mean \gls{twi} $\approx 794$--$879$\,ms, whereas {Uniform} allocation \gls{twi} increases and becomes highly volatile at ${H}_{\max}{=}4$; mean $1713.4$\,ms with a exorbitantly wide min--max band. The mean gap is approximately $21\%$ for ${H}_{\max}{=}1$--$2$ and rises to $\sim49\%$ at ${H}_{\max}{=}4$. Longer chains increase variance and drop mass, making Uniform allocations prone to push some bottleneck $\alpha_k$ close to $1$ and thus inflate the Cantelli margin; Optimal counters it by concentrating budget on the hardest path and bounding that margin. For readability the vertical axis is clipped at $2000$\,ms; at ${H}_{\max}{=}4$ the {maximum} under Uniform often exceeds this limit substantially ($\approx 4\times 10^{4}$) even though the mean remains within range.
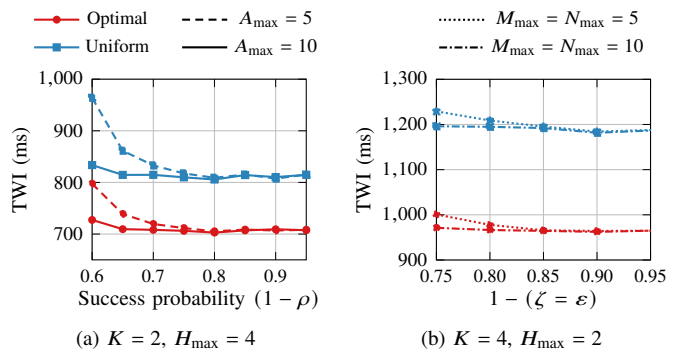
\begin{figure}[b]
    \vspace{-0.7cm}
    \centering
    \subfloat[$K=2$, $H_{\max}=4$]{        
        \begin{tikzpicture}
        \begin{axis}[
            twiLinePlot,
            xlabel={Success probability $(1-\rho)$},
            ylabel={TWI (ms)},
 xmin=0.60, xmax=0.95,
  xtick={0.60,0.7,0.8,0.9},      
  xticklabels={0.6,0.7,0.8,0.9},              
            ymin={650}, ymax={1000}, ytick distance={100},
            legend columns=2, legend style={at={(0.4,1.1)}, anchor=south, /tikz/every even column/.append style={column sep=1em}},transpose legend
        ]
            \addplot [optimalLine] coordinates {
                (0.60,727.3513) (0.65,709.4560) (0.70,708.1451) (0.75,706.2287)
                (0.80,702.8554) (0.85,707.0982) (0.90,709.2203) (0.95,707.3815)
            };
            \addlegendentry{Optimal}

            \addplot [optimalLine,densely dashed,forget plot] coordinates {
                (0.60,797.5215) (0.65,738.8940) (0.70,719.4790) (0.75,711.5483)
                (0.80,705.3084) (0.85,708.6278) (0.90,706.8826) (0.95,708.2341)
            };

            \addplot [uniformLine] coordinates {
                (0.60,833.6313) (0.65,814.5763) (0.70,814.6917) (0.75,809.8319)
                (0.80,805.6935) (0.85,814.2695) (0.90,810.3484) (0.95,814.6998)
            };
            \addlegendentry{Uniform}
            
            \addplot [uniformLine,densely dashed,forget plot] coordinates {
                (0.60,963.9096) (0.65,861.4052) (0.70,832.9690) (0.75,817.4827)
                (0.80,809.0188) (0.85,815.2397) (0.90,807.4429) (0.95,815.0335)
            };

            \addlegendimage{thick,densely dashed}
            \addlegendentry{$A_{\max}=5$}
            
            \addlegendimage{thick}
            \addlegendentry{$A_{\max}=10$}
        \end{axis}
        \end{tikzpicture}
    }
    \subfloat[$K=4$, $H_{\max}=2$]{
        \begin{tikzpicture}
        \begin{axis}[
            twiLinePlot,
            xlabel={$1-(\zeta=\varepsilon)$},
            ylabel={TWI (ms)},
 xmin=0.75, xmax=0.95,
  xtick={0.75,0.80,0.85,0.90, 0.95},      
  xticklabels={0.75,0.80,0.85,0.90, 0.95},              
            ymin={900}, ymax={1300}, ytick distance={100},
            legend columns=1, legend style={at={(0.5,1.1)}, anchor=south, /tikz/every even column/.append style={column sep=1em}},
        ]
            \addplot [optimalLine,densely dashdotted,forget plot] coordinates {
             (0.75, 971.2071) (0.80, 966.3718) (0.85, 964.4652)
                (0.90, 962.2972) (0.97, 965.9576)
            };

            \addplot [optimalLine,densely dotted,forget plot] coordinates {
                (0.75,1000.8398) (0.80, 977.8046) (0.85, 965.7271)
                (0.90, 964.3181) (0.97, 964.9205)
            };

            \addplot [uniformLine,densely dashdotted,forget plot] coordinates {
                (0.75,1196.0183) (0.80,1194.8576) (0.85,1192.0399)
                (0.90,1181.6549) (0.97,1188.5204)
            };
            
            \addplot [uniformLine,densely dotted,forget plot] coordinates {
                (0.75,1229.0741) (0.80,1208.7651) (0.85,1195.3214)
                (0.90,1184.3274) (0.97,1189.0315)
            };

            \addlegendimage{thick,densely dotted}
            \addlegendentry{$M_{\max}=N_{\max}=5$}
            
            \addlegendimage{thick,densely dashdotted}
            \addlegendentry{$M_{\max}=N_{\max}=10$}
        \end{axis}
        \end{tikzpicture}
    }
    \caption{\Gls{twi} obtained by varying (a)~$\rho$ and (b)~$\zeta$.}
    \label{fig:Hop_Succs_Prob_TWI}
\end{figure}
\par Turning to the effect of $T_f$, with $K{=}3$ and ${H}_{\max}{=}2$ held fixed, we observe only a mild upward trend in \gls{twi} across the range $T_f\in\{4,10,16,22\}$\,ms in Fig.~\ref{fig:Varying_Params_TWI}(c): {Optimal} rises from $808.7$ to $852.7$\,ms, while {Uniform} increases from $1028.7$ to $1063.7$\,ms. The dispersion is comparatively stable over this sweep. This weak sensitivity suggests that, under our parameter ranges, the dominant contributors to the bottleneck ({propagation}: $\mu_k$,\,{hop variances}: $\text{Var}(T_2)$,\,{drop masses}: $p_k$) are not strongly modulated by $T_f$; increasing $T_f$ slightly expands per-hop processing/aggregation windows, nudging the means but not changing which path is acting as bottleneck. In contrast, sweeping the cell radius $D$ shows a strong, monotone increase in \gls{twi}, as observed in Fig.~\ref{fig:Varying_Params_TWI}(d). As $D$ grows from $50$ to $200$\,m (again with $K{=}3$, ${H}_{\max}{=}2$), the {Optimal} mean increases from $486.9$ to $1519$\,ms, and {Uniform} from $576.4$ to $1981.5$\,ms. Throughout, {Uniform} remains higher and exhibits a significantly wider min--max envelope than Optimal.
\par Next, in  Fig.~\ref{fig:Hop_Succs_Prob_TWI}(a), we isolate the $T_2$ reliability knob $\rho$ (per-attempt success), with $K{=}2$ and ${H}_{\max}{=}4$ fixed. Increasing $\rho$ systematically lowers \gls{twi} for both policies, and allowing more retries further helps, especially when $\rho$ is modest. For example, at $\rho{=}0.60$ the mean drops from $797.5$ to $727.3$\,ms for Optimal and $963.9$ to $833.6$\,ms for Uniform when moving from $A_{\max}{=}5$ to $10$, reflecting reduced drop mass and a tighter Cantelli margin. As $\rho$ increases toward $0.85$–$0.95$, both curves flatten and the $A_{\max}$ benefit diminishes (e.g., $708.2$ versus $707.2$\,ms at $\rho{=}0.95$ for Optimal), indicating that $T_2$ ceases to be the bottleneck and residual delay/reliability is dominated by other parameters, i.e., \gls{sr} and \gls{pt}. Across all settings Optimal remains consistently below Uniform, owing to its reallocation of the error budget toward the hardest path(s). The results in Fig.~\ref{fig:Hop_Succs_Prob_TWI}(b) for the coupled \gls{sr}/\gls{pt} sweep exhibits the same trend as Fig.~\ref{fig:Hop_Succs_Prob_TWI}(a), owing to them being of the same distribution. 
\begin{figure}[t]
    \centering
    \ref{floating-legend2}\vspace{-5pt}
    \subfloat[$D=100$~m]{        
        \begin{tikzpicture}
        \begin{axis}[
            twiLinePlot,
            ylabel={TWI (ms)},
            x dir=reverse, xmin=0.02, xmax=0.10,
  xtick={0.10,0.08,0.06,0.04,0.02},      
  xticklabels={10,8,6,4,2},              
  xlabel={$\,\varepsilon\ (\cdot 10^{-2})$}, 
            ymin={0}, ymax={8000}, ytick distance={2000},
            legend to name=floating-legend2, legend columns=-1, legend style={/tikz/every even column/.append style={column sep=1em}},
        ]
            \addplot [optimalLine] coordinates {
    (0.10,818.2) (0.09,859.1) (0.08,887.5) (0.07,946.6) (0.06,997.1)
    (0.05,1091.1) (0.04,1223.1) (0.03,1587.3) (0.02,2715.8) (0.01,6185.8)
            };
            \addlegendentry{Optimal}

            \addplot [uniformLine] coordinates {
    (0.10,1044.9) (0.09,1101.2) (0.08,1166.4) (0.07,1711.6) (0.06,1981.1)
    (0.05,2846.0) (0.04,3769.2) (0.03,6055.2) (0.02,12279.2) (0.01,20381.2)
  };
            \addlegendentry{Uniform}

            \addlegendimage{thick}
            \addlegendentry{$D=100$~m}
            
            \addlegendimage{thick,densely dashed}
            \addlegendentry{$D=50$~m}

            \addplot[name path=optimal_uppbound,draw=none] coordinates {
    (0.10,1247.1) (0.09,1390.0) (0.08,1378.2) (0.07,1420.9) (0.06,1989.4)
    (0.05,3191.2) (0.04,16336.2) (0.03,21708.4) (0.02,24798.7) (0.01,30440.9)
  };
            \addplot[name path=optimal_lowbound,draw=none,domain={1:12}] coordinates {
    (0.10,370.5) (0.09,383.7) (0.08,399.9) (0.07,419.3) (0.06,443.4)
    (0.05,473.8) (0.04,515.1) (0.03,575.9) (0.02,677.7) (0.01,906.5)
  };
            \addplot[optimalArea] fill between [of=optimal_uppbound and optimal_lowbound];

            \addplot[name path=uniform_uppbound,draw=none] coordinates {
    (0.10,2640.2) (0.09,5484.7) (0.08,17548.6) (0.07,45157.8) (0.06,45156.6)
    (0.05,45159.0) (0.04,45161.7) (0.03,45165.7) (0.02,45165.5) (0.01,45164.0)
  };
            \addplot[name path=uniform_lowbound,draw=none,domain={1:12}] coordinates {
    (0.10,427.0) (0.09,443.2) (0.08,462.6) (0.07,486.3) (0.06,515.7)
    (0.05,552.9) (0.04,603.4) (0.03,677.9) (0.02,802.1) (0.01,1083.0)
  };
            \addplot[uniformArea] fill between [of=uniform_uppbound and uniform_lowbound];
        \end{axis}
        \end{tikzpicture}
    }
    \subfloat[$D\in\{50,100\}$~m]{
        \begin{tikzpicture}
        \begin{axis}[
            twiLinePlot,
            ylabel={TWI (ms)},
            x dir=reverse, xmin=0.02, xmax=0.10,
      xtick={0.10,0.08,0.06,0.04,0.02},      
      xticklabels={10,8,6,4,2},              
      xlabel={$\,\varepsilon\ (\cdot 10^{-2})$}, 
            ymin={0}, ymax={8000}, ytick distance={2000},
            legend columns=2, legend style={at={(0.4,1.1)}, anchor=south, /tikz/every even column/.append style={column sep=1em}},
        ]
            \addplot [optimalLine,forget plot] coordinates {
    (0.10, 818.2234) (0.09, 859.0899) (0.08, 887.5147) (0.07, 946.6123)
    (0.06, 997.0634) (0.05,1091.0954) (0.04,1223.1056) (0.03,1587.3024)
    (0.02,2715.7559) (0.01,6185.8009)
  };

            \addplot [optimalLine,densely dashed,forget plot] coordinates {
    (0.10, 487.6676) (0.09, 509.1285) (0.08, 531.1172) (0.07, 560.0005)
    (0.06, 596.8738) (0.05, 648.5597) (0.04, 727.7923) (0.03, 952.2780)
    (0.02,1622.9756) (0.01,3696.3583)
  };

            \addplot [uniformLine,forget plot] coordinates {
    (0.10, 1044.8709) (0.09, 1101.1700) (0.08, 1166.3903) (0.07, 1711.5927)
    (0.06, 1981.0734) (0.05, 2845.9772) (0.04, 3769.2359) (0.03, 6055.2189)
    (0.02,12279.1719) (0.01,20381.2446)
  };
            
            \addplot [uniformLine,densely dashed,forget plot] coordinates {
    (0.10,  577.9551) (0.09,  605.3101) (0.08,  657.3277) (0.07,  965.5732)
    (0.06, 1166.2109) (0.05, 1775.5600) (0.04, 2277.3267) (0.03, 3717.6938)
    (0.02, 7569.2430) (0.01,12044.8408)
  };
        \end{axis}
        \end{tikzpicture}
    }
    \caption{\Gls{twi} obtained by varying $\varepsilon$. $K=3$, $H_{\max}=2$.}
    \label{fig:Varepsilon_TWI}
\end{figure}
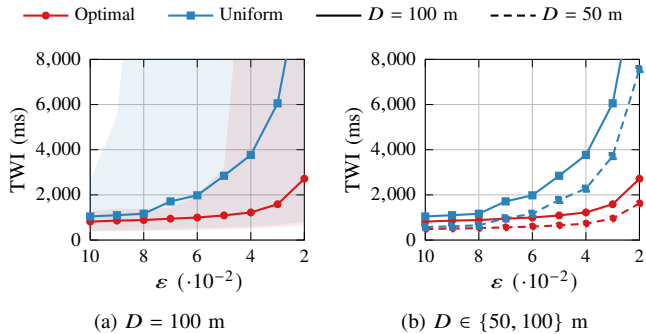
\par Finally, in Fig.~\ref{fig:Varepsilon_TWI}, we analyse the effect of varying the overall reliability budget $\varepsilon$ for $K=3$, ${H}_{\max}=2$, and $D{=}100$. In Fig.~\ref{fig:Varepsilon_TWI}(a), as $\varepsilon$ tightens (smaller $\varepsilon$), the achievable \gls{twi} under Uniform allocation rises sharply and becomes increasingly volatile; the min–max envelope  expands exorbitantly as well. In contrast, the Optimal allocation keeps $\mathrm{TWI}$ comparatively stable by handling harder paths well. Its min–max range also remains tight down to low $\varepsilon$ (approximately $\varepsilon\approx0.03$), and only then grows markedly for extremely small budgets. For a smaller distance $D{=}50$ in Fig.~\ref{fig:Varepsilon_TWI}(b), the same qualitative trend holds but the entire curve shifts down for both. The min-max range is less volatile over $\varepsilon{=}0.10\!\to\!0.01$ for Optimal allocation. Intuitively, shorter paths (smaller $D$) exhibit smaller means/variances and fewer drop-heavy chains, so both policies benefit; yet {Optimal} consistently delivers lower $\mathrm{TWI}$ and tighter dispersion across all reliability  budgets.
\section{Conclusion and Future Works}\label{Conc}
In this work, we formalized \gls{twi}–Causality for perception-oriented timing {in support of Physical \gls{ai}}, with a singular focus on the {usefulness} guarantee. Delivery is gated by a \gls{twi} that specifies when information remains actionable at the network–application interface, while simultaneity and per-path causality are assumed upstream. Within a cellular pipeline comprising sensing/propagation, computation, and access/transmission stages under slotted framing with bounded retransmissions, we developed a tractable end-to-end delay/reliability model per causal path. We then formulated \gls{twi} synthesis as minimizing a common validity horizon subject to a delivery-reliability constraint, jointly allocating per-path violation budgets via a convex program that admitted a globally optimal solution. First, in specific two-path scenarios, we illustrated that optimized (non-uniform) reliability allocation, by favouring the hardest path, yields smaller and more stable horizons than the uniform-after-threshold baseline. Thereafter, in general randomized topologies, the experiments surfaced the dominant drivers of the required horizon, i.e., path depth/width, cell size, frame length, and access timing, thereby providing design guidance for tuning reliability budgets and network parameters {for timing-aware perception over multimodal sensor fields.}
\par This study isolates the usefulness primitive and synthesizes a minimal \gls{twi} under a delivery-reliability constraint. Looking ahead, we envision and outline multiple extensions. First, within the scope of this work, investigate tighter reliability bounds using higher-order moments and beyond–moment characterizations to sharpen timeliness guarantees. Second, building on this work, define simultaneity and its associated grouping policy over the admissible causal paths, followed by joint optimization of the coincidence window and the validity horizon. Third, incorporate causality preservation into the framework, ensuring correct order along paths, which in turn entails joint optimization of network parameters and reliability budgets. In parallel, examine modelling refinements such as bursty access, heavy-tailed retries, and bounded clock error. Ultimately, these extensions should converge to a single coherent design that couples path selection, alignment, ordering, and usefulness under unified reliability and resource constraints. {Together, these extensions will move \gls{twi}–Causality from an optimal sizing tool to a deployable timing discipline that enable timing-aware systems such as Physical \gls{ai} by fusing multimodal updates with alignment, order, and actionability.}
\appendix
\subsection{Bounded Delay Range}\label{BDR_FS}
As delineated in Section \ref{Sys_Mod}, the propagation delay between the event and the sensor is bounded as 
$T_{\text{ES}}^{\text{prop}} \in [0, \frac{2D}{v} ]$. Moreover, the computation delay is modelled as \( C_1 \sim \mathcal{U}[C_{\min}, C_{\max}] \). 
Adding the two independent components, the total delay before ceiling satisfies 
$T_{\text{ES}}^{\text{prop}} + C_1 \in [ C_{\min}, ~ \frac{2D}{v} + C_{\max} ]$.
Dividing by the frame duration \( T_f \), we obtain the bounds:
\begin{equation}
\nonumber
(T_{\text{ES}}^{\text{prop}} + C_1)/T_f \in [n_{\min}, n_{\max} ],
\end{equation}
where $n_{\min}=\frac{C_{\min}}{T_f}$ and $n_{\max}=\frac{2D}{vT_f} + \frac{C_{\max}}{T_f}$, which implies that the ceiling operation in the delay calculation is over a finite range. This justifies computing the expected delay \( \mathbb{E}[T_1] \) via a finite sum:
\begin{equation}
\textstyle
\mathbb{E}[T_1] = \sum_{n=n_{\min}}^{n_{\max}} n T_f \cdot \Pr( \lceil (T_{\text{ES}}^{\text{prop}} + C_1)/T_f \rceil = n ).
\end{equation}
\subsection{Derivation of Sensor-to-Sensor Delay}\label{STS_DM}
We compute the \gls{pmf} and expectation of 
$T_2 = T_f \cdot \lceil a_{k,i} + \frac{C_{k,i}}{T_f} \rceil$ by first conditioning on the discrete variable \( a_{k,i} \in \{1, 2, \dots, A_{\max}\} \), which follows a truncated geometric distribution:
\begin{equation}
\nonumber
\mathbb{P}(a_{k,i} = a) = \rho_{k,i}^{a-1}(1 - \rho_{k,i}) / (1 - \rho_{k,i}^{A_{\max}}).
\end{equation}
For fixed \( a \), the expression \( a + \frac{C_{k,i}}{T_f} \) lies in the interval 
$[ a + \frac{C_{\min}}{T_f}, ~ a + \frac{C_{\max}}{T_f} ]$ and its ceiling takes integer values
$n \in [\lceil a + \frac{C_{\min}}{T_f} \rceil, ~ \lceil a + \frac{C_{\max}}{T_f} \rceil]$.
Thus, the conditional \gls{pmf} of the integer-valued random variable 
$N_{2} = \lceil a + \frac{C_{k,i}}{T_f} \rceil$, denoted as $p_{a,n}=\mathbb{P}(N_{2} = n\,|\,a)$,~is 
\begin{equation}
\nonumber
p_{a,n} = \frac{\min(C_{\max}, T_f(n - a)) - \max(C_{\min}, T_f(n - 1 - a))}{C_{\max} - C_{\min}},
\end{equation}
with the convention that the numerator is zero if the interval is empty. The conditional expectation is then:
\begin{equation}
\nonumber
\textstyle
\mathbb{E}_{C_{k,i}}[\lceil a + C_{k,i}/T_f \rceil] =
\sum_{n = \lceil a + C_{\min}/T_f \rceil}^{\lceil a + C_{\max}/T_f \rceil}
n \cdot p_{a,n}.
\end{equation}
Finally, the total expectation over \( T_2 \) is:
\begin{equation}
\textstyle
\mathbb{E}[T_2] = T_f \cdot \sum_{a=1}^{A_{\max}} 
\mathbb{P}(a_{k,i} = a) \sum_{n = \lceil a + C_{\min}/T_f \rceil}^{\lceil a + C_{\max}/T_f \rceil} 
n \cdot p_{a,n}.
\end{equation}
This expression allows exact computation of \( \mathbb{E}[T_2] \).
\subsection{Variance of $T^{\text{SR}}_{k,i}$ and $T^{\text{PT}}_{k,i}$}\label{Var_TSR_TPT}
Here, for brevity, we only delineate the variance of the \gls{sr} delay, as both $T^{\text{SR}}_{k,i}$ and $T^{\text{PT}}_{k,i}$ are identical distributions.
Let $r=\zeta_{k,i}\in(0,1)$ and $M=M_{\max}$. With \gls{pmf}
\begin{equation}
\nonumber
\Pr(m)=(1-r)\,r^{\,m-1}/(1-r^{M}), \quad m=1,\dots,M,
\end{equation}
and $T^{\text{SR}}_{k,i}=T_f m$, define
$S_1(r,M)=\sum_{m=1}^{M} m\,r^{\,m-1}$ and $S_2(r,M)=\sum_{m=1}^{M} m^{2}\,r^{\,m-1}$.
Then, we have
\begin{equation}
\nonumber
\mathbb{E}[m]=\frac{1-r}{1-r^{M}}\,S_1(r,M),\quad
\mathbb{E}[m^2]=\frac{1-r}{1-r^{M}}\,S_2(r,M),
\end{equation}
and
$\operatorname{Var}\!\big(T^{\text{SR}}_{k,i}\big)
= T_f^{2}\,[
\mathbb{E}[m^2]
-\left(\mathbb{E}[m]\right)^{\!2}]$, 
where the finite sums admit compact expressions:
\begin{align}
\label{eq:S1_closed}
S_1(r,M)&=[1-(M+1)r^{M}+Mr^{M+1}]/(1-r)^2,\\
\label{eq:S2_closed}
S_2(r,M)&\textstyle= r\,\frac{d}{dr}S_1(r,M)+S_1(r,M).
\end{align}
Substituting \eqref{eq:S1_closed}–\eqref{eq:S2_closed} into expression of $\operatorname{Var}\!\big(T^{\text{SR}}_{k,i}\big)$ yields its closed form. Similarly, the variance of $T^{\text{PT}}_{k,i}$ follows identically by setting 
$r=\epsilon_{k,i}$ and $M=N_{\max}$.
\printbibliography
\end{document}